\documentclass[12pt,letterpaper]{article}
\usepackage{setspace}
\usepackage{color}
\usepackage{amsmath}
\usepackage{amsfonts}
\usepackage{verbatim}
\usepackage{amssymb}
\usepackage{bm}
\usepackage{graphics,graphicx,xcolor}
\usepackage{bbm}
\usepackage{cite}
\usepackage{authblk}
\usepackage{amsthm}
\usepackage[shortlabels]{enumitem}
\usepackage{authblk}

\newcommand{\x}{\mathrm{x}}
\newtheorem{theorem}{Theorem}

\numberwithin{equation}{section}

\title{\textbf{Hidden symmetries from distortions of the conformal structure}}

\author{N. Dimakis\footnote{nsdimakis@scu.edu.cn; nsdimakis@gmail.com}}
\affil{Center for Theoretical Physics, College of Physics, Sichuan University, Chengdu 610065, China}

\date{}

\begin{document}

\maketitle

\begin{abstract}
  It is well established that the mass parameter breaks the conformal symmetries in the case of geodesic motion. The proper conformal Killing vectors cease to generate conserved charges when non-null geodesics are considered. We examine how the introduction of the mass is actually related to the appearance of appropriate distortions in the conformal sector, which lead to new conservation laws. As a prominent example we use a general pp-wave metric, which exploits this property to the maximum. We study the necessary geometric conditions, so that such types of distortions are applicable. We show that the relative vectors are generators of disformal transformations and prove their connection to higher order (hidden) symmetries. Except from the pp-wave geometry, we also provide an additional example in the form of the de Sitter metric. Again, the proper conformal Killing vectors can be appropriately distorted to generate conserved quantities for massive geodesics. Subsequently, we proceed by introducing an additional symmetry breaking effect. The latter is realized by considering a Bogoslovsky type of line-element, which involves a Lorentz violating parameter. We utilize once more the pp-wave case as a guide to study how the broken symmetries - this time also related to Killing vectors - are substituted by distortions of the original generators. We further analyze and discuss the necessary geometric conditions that lead to the emergence of these distortions.
\end{abstract}

\tableofcontents

\section{Introduction}

The motion of particles in curved backgrounds is essential for the understanding of various gravitational phenomena. Among these we may distinguish the study of black hole shadows \cite{Synge,Maeda,Mann,Haitang,BHshadCh} or aspects of the gravitational memory effect \cite{Gib1,Gib2,Shore,Chakra}, both of which are related to the investigation of geodesic motion. Of special importance are cases where the geodesic system is characterized by enough existing integrals of motion, in involution, so as to be deemed as integrable in the Liouville sense. It is well known that, for the geodesic systems of equations, at least in the context of Riemannian geometry, the integrals of motion are closely related to the symmetries of the background manifold. For works regarding the symmetries of the geodesic equations as well as their geometrical significance, see \cite{Katzin,Cav,Hoj,Rosquist,Andr1,Andr2,Gomis,PetHar,Andr3}.

As far as space-time vectors are concerned, the homothetic algebra of the metric is particularly important in the generation of symmetries for the affinely parametrized geodesics \cite{Andr1}, giving rise to point symmetry transformations. On the other hand, Killing tensors of various ranks are connected with the generation of what we refer to as higher order, or hidden, or dynamical symmetries. An example of such a symmetry, which happens to be crucial for the integrability of the relative system, is the one associated to the Carter constant for the motion in a Kerr black-hole background \cite{Carter}. The seminal work by Carter motivated further studies on the subject \cite{Penrose,Woodhouse,Frolov}. In classical mechanics we have of course the well-known examples of the Laplace-Runge-Lenz vector for the Kepler problem \cite{Goldstein} and the Fradkin tensor for the three-dimensional isotropic harmonic oscillator \cite{Fradkin}. In more modern concepts, interesting cases of hidden symmetries arise within the scope of supersymmetry \cite{AA1,AA2,M1,M2,M3,M4}. For more results on higher order or hidden symmetries in various systems, and the geometric conditions of the involved objects in their construction, see also \cite{Kalotas,Carhid,Katzin2,PWHiggs,Pal,Ply,Ply2,Ply3,Tsamp1,Tsamp2,Liv3}.

Apart from the Killing vectors or tensors however, there is an intriguing involvement of the conformal algebra of the metric. For example, in the case of null geodesics, all  conformal Killing vectors (CKVs) generate conserved quantities which are linear in the momenta. However, when time-like geodesics are considered, the relative conserved quantities are just generated by the sub-set of the Killing vectors (KVs). We may say that, the introduction of mass for the test particle, leads to a symmetry breaking effect that reduces the dimensionality of the symmetry group since obviously KVs$\subset$CKVs. Interestingly enough, it has been shown \cite{dimgeo}, that the proper conformal Killing vectors, i.e. the elements of the set CKVs$\cap$KVs, are still involved in the construction of conserved charges, even in the massive case. However, what happens now is that they enter in non-local conserved quantities. Moreover, for their derivation, it is important to maintain the initial parametrization invariant form of the problem, e.g. not to start from the affinely parametrized equations, which explains why they are usually overlooked.

Among the various studies regarding geodesic equations, a great number is devoted to the motion in a pp-wave background \cite{pp1,pp2,pp3,pp4,pp5,pp6,pp7,pp8,ppwaves,BFgeo}. The symmetries of the pp-wave metrics have been extensively studied in several works \cite{Goenner,Maartens,Tupper}. The plane-fronted gravitational waves with parallel rays (or more simply pp-waves \cite{Kundt}) are non-flat space-times defined by the existence of a covariantly constant and null bi-vector. In many cases, another definition is encountered in the literature, that of possessing a covariantly constant, null vector \cite{SthephMac}. The latter however is not equivalent to the first; the two become  indistinguishable, if we just regard vacuum solutions of General Relativity. Here, we follow mainly the formalism of \cite{Maartens,Tupper} and make use of the first definition, which is slightly more restrictive: It implies not only the existence of a covariantly constant, null vector, but also that the space-time is either of type $N$ or $O$ in the Petrov classification \cite{Tupper,Steele}.

The pp-waves have several interesting properties. They belong to a larger class of geometries, whose curvature scalars are all zero, the so called Vanishing Scalar Invariants Space-times (VSI) \cite{Coley}. Their metrics contain the very interesting case of plane gravitational waves and - through Penrose's limit \cite{Penlim} - their applications extend even to string theory \cite{Ortin}. In \cite{ppwaves}, it was shown that, when the background metric is that of a pp-wave spacetime, the non-local conserved charges of the general geodesic problem reduce to local expressions. The resulting integrals of motion appear as if they are generated from mass dependent distortions of the proper conformal Killing vectors of the metric, ``reinstating'', in a sense, the broken symmetries due to the introduction of the mass. Here, we further investigate the nature and the geometric implications of such vectors and show that they still emerge even if you consider a Finslerian generalization of the pp-wave geometry, which introduces an additional, Lorentz violating, parameter.

The outline of this work is the following: First, we start with an overview regarding the existence of non-local integrals of motion for massive geodesics in a generic space-time. We prove that these conserved charges are actually generated by non-local Noether symmetries and we derive their generators. We revisit the result, first appeared in \cite{ppwaves}, about the mass-dependent distortions of the proper conformal Killing vectors, which generate integrals of motion in the case of pp-waves. We concentrate on their geometric interpretation and prove that such vectors are the reduced form of higher order Noether symmetries. What is more, we demonstrate that there exist geometries, besides pp-waves, that can admit such types of ``distorted'' symmetries; as a brief example we consider the de Sitter solution of General Relativity. In the subsequent sections we extend the previous results in the case where a Lorentz violating parameter is also introduced, causing an extra symmetry breaking in conjunction to the mass. This is realized by taking the generalized Bogoslovsky-Finsler line-element. We further investigate the necessary geometric conditions for such distorted vectors to exist and we derive the explicit expressions for the Finslerian pp-waves.

\section{Mass distorted symmetry vectors}

In this section, for the convenience of the reader, we revisit some known facts from the theory of geodesic systems and also make a brief review of the results obtained in \cite{dimgeo} and \cite{ppwaves}, parts of which are going to be of importance in our analysis. We briefly describe the notion of non-local conservation laws related to conformal Killing vectors, as introduced in \cite{dimgeo}, for a general geodesic system. We additionally prove that these conserved quantities are owed to non-local Noether symmetries of the action and present the relative expressions. Subsequently, we proceed to revisit the specialization of this result in the case of pp-wave space-times, where the conformal vectors acquire mass dependent distortions in order to generate conserved charges for time-like geodesics \cite{ppwaves}. For a generic space-time, we investigate the geometric implications of such vectors and show that they are related to higher order (hidden) Noether symmetries. Finally, in order to demonstrate that there can be other geometries - beside pp-waves - admitting such type of symmetries, we present an example utilizing the de Sitter metric.

\subsection{Generic geodesic systems and non-local integrals of motion}

For the motion of a relativistic particle of mass $m$ in a background spacetime whose metric is given by $g_{\mu\nu}$, we consider the action
\begin{equation} \label{primact}
  S[\lambda]  = \int\!\! L d\lambda ,
\end{equation}
where
\begin{equation} \label{primLag}
   L = \frac{1}{2 n} g_{\mu\nu}\dot{\x}^\mu\dot{\x}^\nu- \frac{m^2}{2} n.
\end{equation}
The latter is a quadratic, parameterization invariant Lagrangian. The $\lambda$ denotes the parameter along the trajectory, the $\x^\mu=\x^\mu(\lambda)$ are the coordinates and $n=n(\lambda)$ is an auxiliary degree of freedom referred to as the einbein \cite{einbein}. The dot in \eqref{primLag} is used to symbolize the derivatives with respect to $\lambda$, i.e. $\dot{\x}^\mu= \frac{d \x^\mu}{d\lambda}$.

Under an arbitrary change of the parameter $\lambda\mapsto \tilde{\lambda} = f(\lambda)$, and the transformation laws
\begin{equation} \label{ptrlaw}
  n(\lambda) d\lambda \mapsto n(\tilde{\lambda}) d\tilde{\lambda} \quad \text{and} \quad x(\lambda) \mapsto \tilde{x}(\tilde{\lambda}),
\end{equation}
the action \eqref{primact} remains form invariant, i.e. $S[\lambda]=S[\tilde{\lambda}]$. Hence, arbitrary transformations of the parameter $\lambda$ constitute symmetries of $S$. It is for this reason that Lagrangian \eqref{primLag} is referred to as parametrization invariant. We observe from \eqref{ptrlaw} that, although $n$ is considered a degree of freedom on equal footing with the $\x^\mu$, there is the difference that the latter transform as scalars, while $n$ as a density of weight $+1$.

Maybe the most well-known Lagrangian, used to describe the motion of a relativistic massive particle, is the square root Lagrangian
\begin{equation} \label{sqLAg}
  L_{\text{sq}}=- m \sqrt{-g_{\mu\nu}\dot{\x}^\mu \dot{\x}^\nu},
\end{equation}
where we use the minus inside the square root because we adopt the convention $g_{\mu\nu}\dot{\x}^\mu \dot{\x}^\nu<0$ for time-like geodesics (throughout this work we also make use of the units $c=1$). Lagrangian \eqref{sqLAg} is also parametrization invariant, but this time not because of an auxiliary field, like \eqref{primLag}, but because it is a homogeneous function of degree one in the velocities, i.e. $L_{\text{sq}}(\x,\sigma \dot{\x})=\sigma L_{\text{sq}}(\x, \dot{\x})$, where $\sigma$ is a positive constant.

At this point it is useful to remind Euler's theorem on homogeneous functions, which states that: If $h(y)$ is a homogeneous function of degree $k$, i.e. $h(\sigma y) = \sigma^k h(y)$ for $\sigma>0$, then the following equality holds
\begin{equation}\label{Eultheo}
  y^\mu \frac{\partial h}{\partial y^\mu} = k h .
\end{equation}
By simply setting $h=L_\text{sq}$, $y=\dot{\x}$ and $k=1$ in \eqref{Eultheo}, the theorem, in the case of Lagrangian \eqref{sqLAg}, implies that the latter has an identically zero Hamiltonian, $\dot{\x}^\mu\frac{\partial L_{\text{sq}}}{\partial \dot{\x}^\mu}-L_{\text{sq}}\equiv 0$. This, together with the fact that the euler-Lagrange equations of \eqref{sqLAg} are not well-defined for null geodesics (the expression $g_{\mu\nu} \dot{\x}^\mu \dot{\x}^\nu=\dot{\x}^\mu \dot{\x}_\mu=0$ appears in denominators), makes the use of the $L$ of \eqref{primLag} better suited for our purposes.

The einbein Lagrangian of \eqref{primLag} is dynamically equivalent to $L_{\text{sq}}$. In order to see this we need to write down the Euler-Lagrange equations of \eqref{primLag}, which are equivalent to
\begin{subequations}
\begin{align} \label{eulgenLx}
  & \ddot{\x}^\mu + \Gamma^\mu_{\kappa\lambda} \dot{\x}^\kappa \dot{\x}^\lambda  = \dot{\x}^\mu \frac{d}{d\lambda}\left( \ln n\right) \\
  & \frac{1}{n^2} g_{\mu\nu} \dot{\x}^\mu \dot{\x}^\nu+ m^2  =0 , \label{eulgenLn}
\end{align}
\end{subequations}
where the $\Gamma^\mu_{\kappa\lambda}$ are the Christoffel symbols of the metric $g_{\mu\nu}$. The first set consists of the second order equations obtained by variation with respect to $\x$, while the last equation, \eqref{eulgenLn}, is the constraint equation acquired by variation with respect to $n$. By solving algebraically this last relation for the einbein, $n$, and substituting it in equations \eqref{eulgenLx}, we obtain the Euler-Lagrange equations of $L_{\text{sq}}$. In the einbein formalism, the affinely parametrized geodesics are obtained by using the gauge fixing condition $n=$constant, which leads, from \eqref{eulgenLn}, to $\do{\x}^\mu \dot{\x}_\mu=$const. What is more, for the null geodesics, we need to just set $m=0$ in \eqref{eulgenLn}, which leads to no  complications \eqref{eulgenLx}.

Unlike $L_{\text{sq}}$, the Hamiltonian of \eqref{primLag} is not identically zero; it happens to become zero, but in a weak sense according to Dirac's theory of constrained systems \cite{Dirac,Sund}. The total Hamiltonian is obtained through the Dirac-Bergmann algorithm \cite{Dirac,AndBer} and it reads
\begin{equation} \label{Ham}
  H_T = \frac{n}{2} \mathcal{H} + u_n p_n,
\end{equation}
which is a linear combination of constraints $p_n\approx 0$ and $\mathcal{H}\approx 0$. The symbol ``$\approx$'' denotes a weak equality, meaning, that the respective quantities (here $p_n$ and $\mathcal{H}$) cannot be set to zero prior to carrying out Poisson bracket calculations. Only the end result - after calculating Poisson brackets - is meant to be projected on the constraint surface, defined by the equations $p_n=0$ and $\mathcal{H}=0$. The $p_n$ corresponds to the momentum for the degree of freedom $n$ and the relation $p_n\approx 0$ forms the primary constraint of the theory, the $u_n$ is an arbitrary multiplier and
\begin{equation}\label{Hamcon}
  \mathcal{H}= g^{\mu\nu} p_\mu p_\nu + m^2 \approx 0
\end{equation}
is the secondary constraint - also called Hamiltonian or quadratic constraint.  The $p_\mu=\frac{\partial L}{\partial \dot{\x}^\mu}$ are the usual momenta conjugate to the degrees of freedom $\x^\mu$.

As we mentioned, the action \eqref{primact} describes a parametrization invariant system, i.e. one whose action and equations of motion remain invariant under arbitrary changes of the parameter $\lambda$. The symmetry structure of this type of quadratic in the velocities Lagrangians, including a potential term, has been studied in \cite{tchris} together with its connection to minisuperspace cosmological systems in Einstein's General Relativity. For recent studies on the algebra spanned by the symmetries of such a Lagrangian, associated to minisuperspace cosmology, see \cite{Livine1,Livine2}. In particular in what regards geodesic problems, it has been shown \cite{dimgeo}, that the system described by \eqref{primLag} admits non-local conserved quantities of the form
\begin{equation} \label{genericnonlocal}
  I(\lambda,\x,p) = Y^\mu \frac{\partial L}{\partial \dot{\x}^\mu} + m^2 \int\!\! n(\lambda) \omega(\x(\lambda)) d\lambda = Y^\mu p_\mu + m^2 \int\!\! n(\lambda) \omega(\x(\lambda)) d\lambda ,
\end{equation}
where the $Y^\mu$ are the components of conformal Killing vectors of $g_{\mu\nu}$ with conformal factor $\omega(\x)$, i.e.
\begin{equation}\label{confeq}
  \mathcal{L}_Y g_{\mu\nu} = 2 \omega(\x) g_{\mu\nu} ,
\end{equation}
where we use $\mathcal{L}$ do denote the Lie derivative. The charge $I$ has an explicit dependence on the parameter $\lambda$, brought about by the integral we see on the right hand side of \eqref{genericnonlocal}. The total derivative of $I$ with respect to the parameter can be seen that it is zero by virtue of the Hamiltonian constraint:
\begin{equation}
  \frac{dI}{d\lambda} =  \frac{\partial I}{\partial \lambda} + \{I,H_T\} = n \omega(\x) \mathcal{H} \approx 0 .
\end{equation}
The conserved charge given by $I$ in \eqref{genericnonlocal} is non-local due to involving an integral of phase space functions. This means that at least some prior knowledge of the trajectory is in principle needed in order to carry out the integration in the right hand side of \eqref{genericnonlocal} and acquire the explicit dependence on $\lambda$ that $I(\lambda,\x,p)$ has. The parametrization invariance however, can help overcome such a difficulty. To experience this, we need to remember that $n(\lambda)$ can be used to fix appropriately the gauge, i.e. choose the parameter along the curve. For example a choice like $n=\omega(\x)^{-1}$ makes the $I$ for the corresponding conformal Killing vector to become
\begin{equation} \label{Ifixed}
  I = Y^\mu p_\mu + m^2 \lambda .
\end{equation}
Such is the case, when we consider the affinely parametrized geodesics ($n=1$) and a generic homothetic vector ($\omega=1$), which is known to result in an integral of motion like \eqref{Ifixed}, possessing a linear dependence on the parameter $\lambda$. What we see here, with the help of \eqref{genericnonlocal}, is that any proper conformal Killing vector can lead, under the appropriate choice of parameter along the curve, to an integral of motion of the form \eqref{Ifixed}. Thus, we can always ``localize'' at least one of any integrals of motion of the form \eqref{genericnonlocal}, by choosing appropriately the parameter $\lambda$ (a time choice gauge fixing).

We need to mention that, the expression \eqref{genericnonlocal}, also yields two other well-known results from the theory of symmetries of geodesic systems:
\begin{enumerate}[a)]
  \item When $Y$ corresponds to a Killing field, i.e. $\omega(\x)=0$, we obtain the typical conserved quantities of the form $I=Y^\mu p_\mu$.
  \item If we consider null geodesics ($m=0$), then all conformal Killing fields generate conserved quantities of the form $I=Y^\mu p_\mu$.
\end{enumerate}
Obviously, the substitution of either $\omega(\x)=0$ or $m=0$ in \eqref{genericnonlocal} leads to the desired linear in the momenta expressions and thus, we obtain the expected results of the two cases. It is interesting to note, that the two previous properties, signal an explicit symmetry breaking at the level of the Lagrangian \eqref{primLag}. When the parameter $m$ is zero, conformal Killing vectors (CKVs) form symmetries and generate conserved charges. On the other hand, when $m\neq 0$, only pure Killing vectors (KVs) remain with this property. Of course we have KVs$\subset$CKVs, hence, the mass is responsible for breaking a symmetry group. The new information that \eqref{genericnonlocal} provides us with, is that, even when $m\neq0$, the proper conformal Killing fields, still somehow contribute in generating integrals of motion, but of non-local nature. An important question is, if these new charges are actually owed to some Noether symmetries, which substitute the ones broken of the original CKVs. This is what we will prove later, after briefly presenting the concept of Noether symmetries and their charges.

\subsection{Noether Symmetries} \label{secNoether}

We start with a short review of how Noether symmetries are calculated. In this presentation, we use as our model the Lagrangian \eqref{primLag}, since it is the one that it is of interests to us.

If we consider a general transformation in the space of the dependent and independent variables - $n(\lambda)$, $\x^\mu(\lambda)$ and $\lambda$ respectively - then its generator is written as
\begin{equation} \label{upsbN}
X =\chi \frac{\partial}{\partial \lambda}+ X_n \frac{\partial}{\partial n}+ X^\mu \frac{\partial}{\partial \x^\mu},
\end{equation}
where $\chi$, $X_n$ and $X^\mu$ denote the coefficients in the relative directions. If the corresponding transformation leaves form invariant the action \eqref{primact} of the system ($\delta S=0$), we say that it makes up a variational or, more broadly known as, a Noether symmetry transformation. In infinitesimal form, the criterion which tells us if this condition is satisfied reads \cite{Olver}
\begin{equation} \label{generalsymcon}
  \mathrm{pr}^{(1)} X (L) + L \frac{d\chi}{ d\lambda} = \frac{d \Phi}{d \lambda},
\end{equation}
where $\Phi$ is some function related with the surface term up to which the action $S$ may change ($\delta S=0\Rightarrow \delta(Ld\lambda)=d\Phi$) \cite{Sund}. Symmetries that satisfy \eqref{generalsymcon} for $\Phi\neq$const. are sometimes referred to as quasi-symmetries, exactly because they cause the action to change by a surface term. The $\mathrm{pr}^{(1)} X$ is called the first prolongation of the vector $X$. It is the extension of the basic vector $X$ to the space of the first order derivatives $\dot{\x}^\alpha$ and it is given by the formula
\begin{equation} \label{prologform}
  \mathrm{pr}^{(1)} X = X + \left(\frac{dX^\mu}{d\lambda} -\dot{\x}^\mu \frac{d\chi}{d\lambda}\right) \frac{\partial}{\partial \dot{\x}^\mu} .
\end{equation}
We just consider the first prolongation because the Lagrangian $L$ that we use has a dependence up to velocities. For higher order Lagrangians, e.g. containing accelerations, one would also use higher order prolongations.

When a vector satisfies the symmetry criterion \eqref{generalsymcon}, it gives rise to the conserved quantity of the form
\begin{equation}\label{genint}
  I = X^\mu \frac{\partial L}{ \partial \dot{\x}^\mu} + \chi \left(L - \dot{\x}^\mu \frac{\partial L}{\partial \dot{\x}^\mu} \right) - \Phi = X^\mu p_\mu - \chi \mathcal{H} - \Phi,
\end{equation}
where in the last equality we substituted the equivalent phase space expressions for the momenta and the Hamiltonian constraint. In both \eqref{generalsymcon} and \eqref{genint} we neglected terms that would formally appear and have to do with derivatives of the Lagrangian with respect to $\dot{n}$. Since $L$ has no $\dot{n}$ dependence, these terms are bound to be trivially zero.

Up to now, we have made no assumption over the dependencies that the involved functions may have. These classify the symmetry vector, $X$, into different categories. For example if $\chi$, $X_n$ and $X^\mu$ depend only on the independent and dependent variables, $\lambda$, $\x^\mu$ and $n$, then we say that $X$ is a generator of a \emph{point} symmetry. If, on the other hand, there is additional dependence on derivatives, like for example $\dot{\x}^\mu$, then we talk about \emph{higher order} (or \emph{hidden}) symmetries. If there is dependence on non-local expressions, then we refer to $X$ as a \emph{non-local} symmetry generator.

The simplest case is that of point symmetries, because, for them, there exists an algorithmic procedure of deriving the corresponding symmetry generator.  The process is the following: When calculating \eqref{generalsymcon}, we obtain an expression which contains the functions $\chi$, $X_n$, $X^\mu$, $\Phi$ and their derivatives. Due to the presence of $L$ in \eqref{generalsymcon}, there appear terms involving products of velocities $\dot{\x}^\mu$. However, since we consider  a point symmetry, none of the involved functions $\chi$, $X_n$, $X^\mu$ or $\Phi$ depends on velocities. As a result, each coefficient of different velocity products inside \eqref{generalsymcon} needs to be set separately equal to zero. This creates an over-determined system of partial differential equations for the coefficients of the vector $X$ and for $\Phi$, which, when solved, it derives the desirable symmetry vector. For example, in the case of the geodesic Lagrangian \eqref{primLag}, and for $m\neq0$, the Killing vectors of the metric emerge as generators of point symmetries: equation \eqref{generalsymcon} leads to $\mathcal{L}_X g_{\mu\nu}=0$ for $X=X^\mu(\x) \frac{\partial}{\partial \x^\mu}$ and $\Phi=$const. According to \eqref{genint}, the corresponding conserved charge is linear in the momenta, $I=X^\mu p_\mu$.

The situation gets severely more complicated for higher order symmetries. Imagine for example that we allow dependencies on the velocities, $\dot{\x}^\mu$, inside $\chi$, $X_n$, $X^\mu$ and $\Phi$. Then, we cannot proceed in the same manner as before, by breaking equation \eqref{generalsymcon} in smaller pieces according to the different velocity dependencies. The \eqref{generalsymcon} is to be solved in its totality as a single equation. This complexity is what makes higher order symmetries, sometimes to be referred to, as hidden symmetries. In order to facilitate the procedure of encountering such symmetries however, certain restrictions are usually assumed in the dependencies of the velocities inside the aforementioned functions, e.g. consider only polynomial dependencies up to certain order. The most usual case is, when considering a linear dependence in the velocities inside the coefficients of $X$. Then, you obtain integrals of motion associated with Killing tensors of second rank leading, through \eqref{genint}, to quadratic in the momenta constants of the motion. Such is the case of the famous Carter constant in the Kerr geometry, which is related to the existence of a non-trivial Killing tensor $K_{\mu\nu}$. Now, \eqref{generalsymcon} results in a symmetry generator of the form $X=K^\mu_{\phantom{\mu}\nu}(\x) \dot{\x}^\nu \frac{\partial}{\partial \x^\mu}$, under the condition $\nabla_{(\kappa}K_{\mu\nu)}=0$ and $\Phi=$const. Here, $\nabla$ is the covariant derivative and the parenthesis in the indices denotes the usual full symmetrization, e.g. $A_{(\mu\nu)}=\frac{1}{2} \left(A_{\mu\nu}+A_{\nu\mu} \right)$. In this case, \eqref{genint} yields a quadratic in the momenta conserved charge, $I=K^{\mu\nu} p_\mu p_\nu$.

Let us consider the integral of motion \eqref{genericnonlocal}, which is a non-local expression. It is logical to assume that there might be some non-local symmetry generator \eqref{upsbN} satisfying \eqref{generalsymcon} for the einbein Lagrangian \eqref{primLag}. Truly, it is not very difficult to see that if we write the vector
\begin{equation} \label{symgennl}
  X = \left(Y^\mu - \frac{\dot{\x}^\mu}{n}  \int\!\! n \omega(\x) d\lambda \right)\frac{\partial}{\partial \x^\mu} ,
\end{equation}
where $Y$ is a conformal Killing vector satisfying \eqref{confeq}, then this $X$ satisfies \eqref{generalsymcon} for $\Phi=$const. According to the prolongation formula \eqref{prologform}, we obtain for the vector \eqref{symgennl}
\begin{equation}
  \begin{split}
   \mathrm{pr}^{(1)}X & = \left(Y^\mu - \frac{\dot{\x}^\mu}{n}  \int\!\! n \omega d\lambda \right)\frac{\partial}{\partial \x^\mu} + \left( \frac{d Y^\mu}{d \lambda} + \left(\frac{\dot{n} \dot{\x}^\mu}{n^2} -\frac{\ddot{\x}^\mu}{n}  \right)\int\!\! n \omega d\lambda - \omega \dot{\x}^\mu \right)\frac{\partial}{\partial \dot{\x}^\mu} \\
   & = \left(Y^\mu - \frac{\dot{\x}^\mu}{n}  \int\!\! n \omega d\lambda \right)\frac{\partial}{\partial \x^\mu} + \left( \frac{\partial Y^\mu}{\partial \x^\kappa} \dot{\x}^\kappa + \frac{1}{n}\Gamma^{\mu}_{\kappa\lambda} \dot{\x}^\kappa \dot{\x}^\lambda \int\!\! n \omega d\lambda - \omega \dot{\x}^\mu \right) \frac{\partial}{\partial \dot{\x}^\mu}.
  \end{split}
\end{equation}
In the above expression we used the chain rule in order to write $\frac{d Y^\mu}{d \lambda}= \frac{\partial Y^\mu}{\partial \x^\kappa} \dot{\x}^\kappa$ and the equations of motion \eqref{eulgenLx} to eliminate the accelerations $\ddot{\x}^\mu$. The action of the above prolonged vector on the Lagrangian \eqref{primLag} yields
\begin{equation} \label{XactonL}
  \begin{split}
    \mathrm{pr}^{(1)}X (L)  = \frac{1}{2n} \left(\mathcal{L}_Y g_{\mu\nu}-2\omega(\x) g_{\mu\nu}\right) \dot{\x}^\mu\dot{\x}^\nu - \frac{1}{2 n^2} \left(\int\!\! n(\lambda) \omega(\x(\lambda)) d\lambda \right) \nabla_\kappa g_{\mu\nu} \dot{\x}^\mu \dot{\x}^\nu \dot{\x}^\kappa  = 0 .
  \end{split}
\end{equation}
The first term after the equality in \eqref{XactonL} is zero because, by our assumption, $Y$ is a conformal Killing vector satisfying \eqref{confeq}, while  the second also vanishes due to the covariant derivative of the metric being zero, $\nabla_\kappa g_{\mu\nu}=0$. Hence, criterion \eqref{generalsymcon} is satisfied for \eqref{symgennl} with $\Phi=$const. We have thus proved that there exists a non-local symmetry generator \eqref{symgennl}, which gives rise to the integral of motion \eqref{genericnonlocal}. We may now proceed and see how all these apply in the case of a pp-wave geometry and why it is, in a sense, special.

\subsection{The exceptional pp-wave case} \label{secRpp}

A generic pp-wave space-time, in Brinkmann coordinates, is described by a line-element of the form
\begin{equation} \label{lineel}
  ds^2 = g_{\mu\nu}d\x^\mu d\x^\nu= H(u,x,y) du^2 + 2 du dv + dx^2+ dy^2,
\end{equation}
where $H=H(u,x,y)$ is the profile function and $\x^\mu=(u,v,x,y)$ are the coordinates. The expressions for a generic conformal Killing vector \eqref{confeq} and the corresponding conformal factor are well known for pp-wave space-times and in these coordinates are given by \cite{Maartens}:
\begin{subequations}
\label{zetackv}
\begin{align}
  & Y^u  = \frac{\mu}{2}  \delta_{ij} x^i x^j + a_i(u) x^i + a(u), \\ \label{zeta2}
  & Y^v  = - \mu v^2 + \left( x^i a_i'(u)+ 2 \bar{b}(u) - a'(u)\right) v + M(u,x,y) \quad , \quad  i,j=1,2\\
  & Y^{i} = - \left( \mu x^i + a_i\right) v + \gamma_{ijkl} a_j'(u)x^k x^l + \bar{b}(u) x^i - \epsilon_{ij}c(u)x^j  + c_i(u),
\end{align}
\end{subequations}
and
\begin{equation} \label{omega}
  \omega = \omega(u,v,x^i)= \bar{b}(u) + x^i a_i'(u) - \mu v
\end{equation}
respectively. The $a$, $\bar{b}$, $c$, $a_i$, and $c_i$, where $i=1,2$, are all functions of the variable $u$, while $\mu$ is a constant parameter. The function $M(u,x,y)$ needs to satisfy certain integrability conditions, given in the appendix \ref{app0}, while the rest of the functions are connected to the profile $H(u,x,y)$ of the pp-wave through
\begin{equation}
\label{rulH}
 \left[ \mu x^i +a_i(u)\right] \partial_i H = 2 \mu H + 2 a_i''(u) x^i -2 a''(u)+4 \bar{b}'(u) .
\end{equation}
In our relations we use the indices $i,j,k,l$ to denote the coordinates on the two-dimensional flat plane $x^i=(x,y)$. The $\delta_{ij}$ is used as a metric in this surface and we won't bother with distinguishing between upper and lower indices in that plane. For the other two coordinates of $\x^\mu$, namely $u$ and $v$, we use the relative letter as a superscript, when we want to denote the component corresponding in that direction. For example, the $Y^u$ denotes the component of the vector $Y$ in the direction $u$. The symbols like $\partial_u$, $\partial_v$ and $\partial_i$, are used to express in a compact form the relative partial derivatives with respect to the corresponding coordinate, e.g. $\partial_u= \frac{\partial}{\partial u}$, $\partial_i= \frac{\partial}{\partial x^i}$.

If we use the pp-wave space-time metric in \eqref{primLag} we obtain the geodesic Lagrangian
\begin{equation} \label{LagppR}
  L = \frac{1}{2 n} \left(H \dot{u}^2 +2 \dot{u}\dot{v} + \dot{x}^2 + \dot{y}^2 \right) - n \frac{m^2}{2} .
\end{equation}
The Euler-Lagrange equations of the system lead to
\begin{subequations}
\label{eulgen}
\begin{align}
\label{Neul}
E_n(L) := \frac{\partial L}{\partial n} - \frac{d}{d\lambda} \left( \frac{\partial L}{\partial \dot{n}} \right)=0 & \Rightarrow  H(u,x,y) \dot{u}^2+ 2 \dot{u}\dot{v} + \delta_{ij}\dot{x}^i \dot{x}^j + n^2 m^2 =0
\\
\label{ueul}
E_u(L) := \frac{\partial L}{\partial u} - \frac{d}{d\lambda} \left( \frac{\partial L}{\partial \dot{u}} \right)=0 & \Rightarrow \ddot{v} +\partial_i H(u,x,y) \dot{x}^i \dot{u}+\frac{1}{2} \partial_u H(u,x,y)\dot{u}^2 -\frac{\dot{n}}{n}\dot{v}=0 , \\
\label{veul}
E_v(L) := \frac{\partial L}{\partial v} - \frac{d}{d\lambda} \left( \frac{\partial L}{\partial \dot{v}} \right)=0 & \Rightarrow \ddot{u} - \frac{\dot{n}}{n}\dot{u} =0,
\\
\label{xeul}
E_i(L) := \frac{\partial L}{\partial x^i} - \frac{d}{d\lambda} \left( \frac{\partial L}{\partial \dot{x}^i} \right)=0 & \Rightarrow \ddot{x}^i -\frac{1}{2}\partial_i H(u,x,y)\dot{u}^2 -\frac{\dot{n}}{n}\dot{x}^i,
= 0
\end{align}
\end{subequations}
where $E_n$, $E_\mu$ denote the Euler derivatives with respect to $n$ and $\x^\mu=(u,v,x^i)$.

According to what we saw in the previous section, we expect the Killing vectors of the pp-wave metric to be associated with point symmetries of the Lagrangian \eqref{LagppR}, yielding linear in the momenta integrals of motion. The proper conformal Killing vectors are also to be involved, but generally in non-local expressions.

Let us note that, the existence of the covariantly constant null Killing vector field, $\ell=\partial_v$, for any pp-wave metric \eqref{lineel} guarantees the conservation of the momentum $p_v=\frac{\partial L}{\partial \dot{v}}$, whose on mass shell value we symbolize with $\pi_v$; this, in order to distinguish it from the phase space variable $p_v$. In other words, on mass shell we have $p_v=\pi_v=$const., due to the conservation law $\frac{dp_v}{d\lambda}=0$. This implies
\begin{equation} \label{intpv}
  p_v = \pi_v \Rightarrow \frac{\dot{u}}{n} = \pi_v  \Rightarrow n = \frac{\dot{u}}{\pi_v } ,
\end{equation}
which is also the solution to the Euler-Lagrange equation \eqref{veul}. Hence, the auxiliary degree of freedom $n$ is proportional to the velocity $\dot{u}$. Note that this is not a gauge fixing condition for $n$, it is bound to hold for any possible parameterization. By using \eqref{intpv} and \eqref{Neul}, it was shown in \cite{ppwaves} that the generic conformal factor $\omega$ of \eqref{omega} can be written in such a way so as to have
\begin{equation} \label{nomega}
  n\, \omega = \frac{d}{d\lambda}\left(g _{\mu\nu}f^\mu \frac{\dot{\x}^\nu}{\dot{u}} \right) = \frac{1}{\pi_v^2}\frac{d}{d\lambda}\left(g _{\mu\nu}f^\mu \frac{\dot{\x}^\nu}{n} \right) = \frac{1}{\pi_v^2}\frac{d}{d\lambda}\left(f^\mu p_\mu \right),
\end{equation}
where $p_\mu = \frac{\partial L}{\partial \dot{\x}^\mu}= \frac{1}{n}g_{\mu\nu}\dot{\x}^\nu$ are the momenta, and $f$ is a spacetime vector with components
\begin{subequations}
\label{coefef}
\begin{align}
  f^u  = & 0,
 \\ \label{coefefv}
  f^v  = & \frac{1}{2} u  \left(x^i a_i'(u) - a' (u)+2 \bar{b}(u)-2 \mu v \right) + \frac{1}{2} x^i a_i (u) + \frac{\mu}{4} \delta_{ij} x^i x^j
 + \frac{1}{2} a(u) - \frac{m^2}{\pi_v^2} \frac{\mu}{4} u^2\,,
  \\
f^{i} = & -\frac{1}{2}u \left(\mu\, x^i + a_i(u) \right)\,.
\end{align}
\end{subequations}
As a result the generally non-local conserved charge \eqref{genericnonlocal} is expressed in phase space, by virtue of \eqref{nomega}, as
\begin{equation}\label{redintR}
  I = \left(Y^\mu + \frac{m^2}{\pi_v^2} f^\mu\right) p_\mu = \Upsilon^\mu p_\mu
\end{equation}
with $Y$ being a conformal Killing vector and where we introduced a new vector $\Upsilon$ with components
\begin{equation} \label{upsilonR}
  \Upsilon^\mu = Y^\mu + \frac{m^2}{\pi_v^2} f^\mu .
\end{equation}
This vector expresses a mass dependent distortion of the proper conformal Killing vectors $Y$. It can be seen that the contribution of $f$ in $I$ is relevant only when $Y$ is a proper CKV. That is, the pure Killing vectors still generate the known conserved expressions $Y^\mu p_\mu$. It is only when $Y$ is a proper CKV that a mass dependent modification is needed in order to have a conserved quantity.

The corresponding conservation law reads:
\begin{equation} \label{conslawR}
  \frac{dI}{d\lambda} = -2n \Omega E_n(L) - \Upsilon^\mu E_\mu(L) - \frac{m^2}{n} \Omega  \left(n^2-\frac{\dot{u}^2}{\pi_v^2}\right)
\end{equation}
where $\Omega =  \omega - \frac{m^2}{2\pi_v^2} \mu\, u$. The right hand side is zero because of the Euler-Lagrange equations \eqref{eulgen} and the known first integral \eqref{intpv}. The two first terms in the right hand side of \eqref{conslawR} is the result of what you get when you take the total derivative of a typical Noether charge; a linear combination of the Euler-Lagrange equations. The existence of the last term in \eqref{conslawR} however, is something different. It is not an equation of motion, but a first order relation, which is zero due to an already known conserved charge. In other words, relation \eqref{conslawR} gives us a conservation law which holds due to the given constant value of another known integral of the motion. In the next section we are going to study what is the exact relation of the vector $\Upsilon$ in \eqref{upsilonR} and the conserved charge $I$ of \eqref{redintR} with the Noether symmetries of this system.

\subsection{Relation to a Noether symmetry}

In section \ref{secNoether}, we gave a brief description of the typical Noether symmetry approach. As can be seen by \eqref{genint}, linear in the momenta integrals of motion of the form $I=X^\mu p_\mu$ are given by point symmetry generators, i.e. vectors whose components depend purely on the dependent and independent variables (no higher order or non-local dependence):
\begin{equation} \label{upsbN2}
X = X^\mu(\lambda,n,\x) \frac{\partial}{\partial \x^\mu}.
\end{equation}
By utilizing the symmetry criterion \eqref{generalsymcon} it is easy to derive that, for the pp-wave space-time, as for any metric, only the Killing vectors of the space-time generate point symmetries of this form. In particular, for massive geodesics $m\neq 0$, we get that $X$ is a symmetry if $\mathcal{L}_X g_{\mu\nu}=0$. On the other hand, as we saw in the previous section, we were able to write the conserved quantity appearing in \eqref{redintR}, which is a linear in the momenta integral of the motion, but which is generated by a mass dependent distortion of the conformal Killing vectors of the metric, the vector $\Upsilon$ in \eqref{upsilonR}. The latter appears to generate a linear conserved charge even though it is not a Noether symmetry. Let us mention here that conserved quantities are not all necessarily of Noetherian origin. However, in this case, we shall demonstrate that there is actually a relation of $\Upsilon$ to a formal Noether symmetry.

In order to reveal the true Noether symmetry it is enough to naively substitute the constant ratio $\frac{m^2}{\pi_v^2}$ that we see in \eqref{upsilonR}\footnote{Note that there is an extra $\frac{m^2}{\pi_v^2}$ term inside the $f^v$ component of $f$, see relation \eqref{coefefv}, which also needs to be substituted.} with its dynamical equivalent. In other words lets us substitute $m^2$ from \eqref{eulgenLn}, with respect to velocities, and $\pi_v=p_v= \frac{\partial L}{\partial \dot{v}}=\frac{\dot{u}}{n}$, then we obtain
\begin{equation}\label{consttodyn}
  \frac{m^2}{\pi_v^2} = - \frac{g_{\mu\nu} \dot{\x}^\mu \dot{\x}^\nu}{\dot{u}^2}.
\end{equation}
We may now write a new vector $\tilde{\Upsilon}$ whose components are defined as
\begin{equation} \label{Uhid}
  \tilde{\Upsilon}^\mu := \Upsilon^\mu|_{(m,\pi_v)\rightarrow \dot{\x}} = Y^\mu - \frac{g_{\mu\nu} \dot{\x}^\mu \dot{\x}^\nu}{\dot{u}^2} \tilde{f}^\mu,
\end{equation}
where $\tilde{f}:=f|_{(m,\pi_v)\rightarrow \dot{\x}}$. Let us consider the first prolongation of this vector, with the help of formula \eqref{prologform}, in order to extend it in the space of the velocities
\begin{equation}
  \mathrm{pr}^{(1)} \tilde{\Upsilon} = \tilde{\Upsilon}^\mu \frac{\partial}{\partial \x^\mu} +  \dot{\tilde{\Upsilon}}^\mu \frac{\partial}{\partial \dot{\x}^\mu} = \left(Y^\mu - \frac{g_{\mu\nu} \dot{\x}^\mu \dot{\x}^\nu}{\dot{u}^2} \tilde{f}^\mu\right) \frac{\partial}{\partial \x^\mu} + \left(\dot{Y}^\mu - \frac{g_{\mu\nu} \dot{\x}^\mu \dot{\x}^\nu}{\dot{u}^2} \dot{\tilde{f}}^\mu\right) \frac{\partial}{\partial \dot{\x}^\mu} .
\end{equation}
The second equality in the above relation holds on mass shell. The components $\dot{\tilde{\Upsilon}}^\mu$ in general contain accelerations, which however can be eliminated by using the Euler-Lagrange equations \eqref{ueul}-\eqref{xeul} or equivalently by remembering that the ratio containing velocities in \eqref{Uhid} is an on mass shell constant.

It is easy to verify that $\mathrm{pr}^{(1)} \tilde{\Upsilon} (L)=0$, which means that the vector $\tilde{\Upsilon}$ is a Noether symmetry of the action. However, it is not a point symmetry, since its components in \eqref{Uhid} contain dependence on the first derivatives. The vector $\tilde{\Upsilon}$ constitutes a higher order or a hidden symmetry. The corresponding conserved charge, $\tilde{I}$, which is generated by symmetry \eqref{Uhid}, is connected to the $I$ of \eqref{redintR} in the same manner that the $\tilde{\Upsilon}$ is connected to the $\Upsilon$
\begin{equation} \label{genQ}
  \tilde{I} := I|_{(m,\pi_v)\rightarrow p} = Y^\alpha p_\alpha - \frac{g^{\alpha\beta} (f|_{(m,\pi_v)\rightarrow p})^\gamma p_\alpha p_\beta p_\gamma}{K^{\mu\nu} p_\mu p_\nu}.
\end{equation}
In the above relation we have substituted the constant ratio \eqref{consttodyn} with respect to the momenta, as $\frac{m^2}{\pi_v^2} = - \frac{g^{\mu\nu} p_\mu p_\nu}{p_v^2}$ and we have used the trivial second rank Killing tensor $K= \ell \otimes \ell =\partial_v \otimes \partial_v$, which is constructed out of the covariantly constant Killing vector $\ell$. It is clear that $\pi_v^2 = p_v^2 = K^{\mu\nu}p_\mu p_\nu$. The total derivative of $\tilde{I}$ with respect to the parameter $\lambda$, is zero purely by virtue of the Euler-Lagrange equations \eqref{eulgen}.

We thus have, a higher order symmetry generator $\tilde{\Upsilon}$ whose components are given in \eqref{Uhid}. This generates a Noether charge, $\tilde{I}$, that is a rational function in the momenta. The interesting coincidence is that, on mass shell, part of this ratio is already constant, equal to   $\frac{m^2}{\pi_v^2}$. This leads to the reduced expression of the original conserved quantity, which we denoted with $I$, and which has a linear dependence on the momenta. This reduced charge seems as if generated by a mass dependent distortion of the conformal Killing vectors of the metric; the vector $\Upsilon$ with its components supplied by \eqref{upsilonR}. The latter, even though it is not a formal Noether symmetry, has some interesting geometrical implications that offer a generalization of what we see happening in pp-waves.

\subsection{Geometric interpretation and generalizations}

It is interesting to study, whether this nice coincidence that we encountered in the case of pp-wave space-times, where a higher order symmetry of the geodesics is revealed as a mass dependent distortion of the conformal Killing vectors, can be generalized to include other geometries. We shall see that in principle this is possible, in fact let us first state that:
\begin{theorem} \label{theo1}
  For a given manifold with metric $g_{\mu\nu}$, which admits a second rank Killing tensor $K_{\mu\nu}$, any space-time vector $\Upsilon$ satisfying
  \begin{equation} \label{geomups}
    \mathcal{L}_\Upsilon g_{\mu\nu} = 2 \Omega(\x) \left( g_{\mu\nu} + \frac{m^2}{\kappa} K_{\mu\nu} \right),
  \end{equation}
  produces a linear in the momenta conserved charge $I = \Upsilon^\mu p_\mu$ for the corresponding geodesic system by virtue of the Hamiltonian constraint \eqref{Hamcon} and the conserved charge $K^{\mu\nu}p_\mu p_\nu= \kappa$.
\end{theorem}

The proof can be easily deduced by simply taking the Poisson bracket of $I$ with the Hamiltonian constraint \eqref{Hamcon}, which plays the principal role in the time evolution:
\begin{equation}
  \begin{split}
    \{I,\mathcal{H}\} & = \{\Upsilon^\alpha p_\alpha,g^{\mu\nu}p_\mu p_\nu+m^2\}=-\left(\mathcal{L}_{\Upsilon}g^{\mu\nu}\right)p_\mu p_\nu  = 2 \Omega(\x) \left( g^{\mu\nu} + \frac{m^2}{\kappa} K^{\mu\nu} \right)p_\mu p_\nu \\
    & = 2\Omega(\x) \left( g^{\mu\nu} p_\mu p_\nu + m^2 \right) = 2 \Omega(\x) \mathcal{H} \approx 0,
  \end{split}
\end{equation}
with the second equation being valid due to having $K^{\mu\nu}p_\mu p_\nu = \kappa$. That is, the integral of motion $I$ is related to the constant value of the known quadratic integral. In the pp-wave case, we had $\Omega = \omega- \frac{m^2}{2\kappa} \mu u$, where $\omega$ is the conformal factor ($\mathcal{L}_Y g_{\mu\nu}=2 \omega g_{\mu\nu}$), $\Upsilon$ given by \eqref{upsilonR}, $K=\ell\otimes\ell$ and $\kappa=\pi_v^2$.

In addition to the above, we can prove the following:
\begin{theorem} \label{theo2}
  If a space-time with metric $g_{\mu\nu}$, admits a second rank Killing tensor $K(\neq g)$ and a vector $\Upsilon=\Upsilon(\x,\frac{m^2}{\kappa})$ satisfying \eqref{geomups}, then the
  \begin{equation} \label{geomupshigh}
    \tilde{\Upsilon} = \Upsilon(\x,\frac{-g^{\mu\nu} \dot{\x}^\mu \dot{\x}^\nu}{K_{\alpha\beta}\dot{\x}^\alpha \dot{\x}^\beta}),
  \end{equation}
  is a higher order Noether symmetry generator of the geodesic action, yielding the conserved charge of the form
  \begin{equation} \label{geomtIhigh}
    \tilde{I} = \tilde{\Upsilon}^\alpha p_\alpha .
  \end{equation}
\end{theorem}
The proof is quite straightforward to derive and makes use of the fact that $K$ is a Killing tensor; it does not necessarily require that the space-time is a pp-wave. It can be found in detail in the appendix \ref{App1}.

In the pp-wave case we saw that the vectors satisfying \eqref{geomups} obtain the nice form
\begin{equation} \label{splitU}
  \Upsilon = Y + \frac{m^2}{\kappa} f.
\end{equation}
An important observation is, that the vectors $\Upsilon$ do not necessarily close an algebra. In principle this seems counter-intuitive from our usual experience, but it becomes better understood if we think of our situation in terms of the Poisson bracket formalism. Remember that the linear $I$ of the \eqref{redintR} are the on mass shell reduced expressions of the actual charges $\tilde{I}$ of \eqref{genQ}. The Poisson bracket of two reduced charges $I$ will not necessarily give something which happens to also be a reduced expression of some higher order charge. It is the Poisson brackets between two $\tilde{I}$ charges that are bound to be conserved, not those involving the $I$. The vectors $\Upsilon$ are not the actual symmetries, they offer a convenient reduction scheme that allows for simpler calculations and to reveal the higher order - true Noether - symmetries $\tilde{\Upsilon}$, which would be quite more difficult to derive in the conventional manner.

The relation \eqref{geomups} satisfied by $\Upsilon$, reveals the latter as a generator of disformal transformations. These form a generalization of conformal transformations and where initially introduced by Bekenstein \cite{Bekenstein}. Usually, a disformal transformation of the metric is written as
\begin{equation} \label{disfmet}
  \hat{g}_{\mu\nu} = A(x) g_{\mu\nu} + B(x) \ell_\mu \ell_\nu .
\end{equation}
where  $\hat{g}_{\mu\nu}$ is a new ``physical" metric, $\ell_\mu$ is the gradient of some scalar field, i.e. $\ell_\mu=\nabla_\mu \phi$, and $A(x)$, $B(x)$ are scalar functions of the space-time \cite{Lobo}. One motivation behind the introduction of disformal transformations, was to connect different scalar-tensor theories \cite{BenAchour}. For further uses and applications of disformal transformations see \cite{disf1,disf2,disf3,disf4,disf5,disf6}.  We may generalize \eqref{disfmet}, by defining a transformation of the form $\hat{g}_{\mu\nu} = A(x) g_{\mu\nu} + B(x) K_{\mu\nu}$, with $K$ any second rank tensor, which would be compatible with \eqref{geomups}.  In the pp-wave case, a vector like $\Upsilon$, satisfying a relation like \eqref{geomups} for $K^{\mu\nu}=\ell^\mu \ell^\nu$ with $\ell=\partial_v$ a null, Killing vector, is referred as a \emph{null-like disformal Killing vector} in the terminology of \cite{Lobo}. The $\ell$ can be also written as the gradient of some scalar field $\ell_\mu=\nabla_\mu \phi$, where $\phi=u$. Thus, in the pp-wave case, the vectors $\Upsilon$ generate disformal transformations in accordance with definition \eqref{disfmet}. We need to note that, in the original definition \cite{Bekenstein}, the functions $A$ and $B$ depended only on the scalar field $\phi$ and the inner product $\ell^\mu \ell_\mu = \nabla_\mu \phi \nabla^\mu \phi$.  Obviously this is more restrictive than requiring $A$ and $B$ to be space-time functions.

The existence of a vector $\Upsilon$ satisfying \eqref{geomups} signifies that, in order for these conserved charges to appear, there must exist a coordinate transformation, which at the same time is a disformal transformation of the metric involving a Killing tensor $K$. We may proceed to examine an example of a different metric admitting such symmetries.

\subsubsection{The de Sitter example}

As we demonstrated, for pp-waves, you can satisfy equation \eqref{geomups} by distorting appropriately the conformal Killing vectors of the metric. This raises the question, whether there exist other space-times which also have this property and for which we can derive hidden symmetries through \eqref{geomups}. One obvious answer is the flat space, since all the relations that we used for pp-waves can lead trivially to the Minkowski space (in light-cone coordinates) by simply setting the profile function, $H(u,x,y)$, in the line-element \eqref{lineel}, equal to zero. Here, we report another example in the form of the de Sitter universe corresponding to a spatially flat Friedmann--Lema\^{\i}tre--Robertson--Walker (FLRW) space-time that solves Einstein's equations with a cosmological constant.

If we write the line-element in Cartesian coordinates $\x = (t,x,y,z)$ we have
\begin{equation}\label{deSitterlinel}
  ds^2 = - dt^2 +e^{l t} \left( dx^2 +dy^2 +dz^2 \right),
\end{equation}
where $l$ denotes the constant associated with the value of the Ricci scalar, $R=3l^2$ and the cosmological constant $\Lambda=\frac{3 l^2}{4}$, for which the metric \eqref{deSitterlinel} satisfies Einstein's equations $R_{\mu\nu} - \frac{1}{2} g_{\mu\nu} R +\Lambda g_{\mu\nu}=0$.

The Lagrangian that reproduces the geodesic equations is given by
\begin{equation}\label{DSLag}
  L = \frac{1}{2 n} \left[ - \dot{t}^2 + e^{l t} \left( \dot{x}^2 + \dot{y}^2 + \dot{z}^2 \right) \right] - n \frac{m^2}{2}
\end{equation}
and the equations themselves are equivalent to
\begin{subequations} \label{DSeq}
\begin{align} \label{DSeq0}
  & \dot{t}^2 - e^{l t} \left(\dot{x}^2+\dot{y}^2+\dot{z}^2\right)-m^2 n^2 =0 \\ \label{DSeq1}
  & \ddot{t} = \frac{\dot{n} \dot{t}}{n}-\frac{1}{2} l e^{l t} \left(\dot{x}^2+\dot{y}^2+\dot{z}^2\right) \\ \label{DSeq2}
  & \ddot{x}^i= \frac{\dot{x}^i \left(\dot{n}-n l \dot{t}\right)}{n},
\end{align}
\end{subequations}
where $t(\lambda)$, $x^i(\lambda)$ and $n(\lambda)$ are all functions of $\lambda$, which symbolizes the parameter along the curve.

The manifold where the motion takes place is maximally symmetric, thus possessing ten Killing vectors. They of course generate the corresponding linear in the momenta conserved charges of the geodesic equations. We refrain from giving their expressions here. We are more interested in the five proper conformal Killing vectors of the space-time, which, in these coordinates, are written as
\begin{equation} \label{DSCKV}
  \begin{split}
  & Y_0 = e^{\frac{l}{2} t} \partial_t, \quad Y_i = e^{\frac{l}{2} t} x^i \partial_t - \frac{2}{l} e^{-\frac{l}{2} t} \partial_i \\
  & Y_4 = e^{-\frac{l}{2} t} \left(\frac{4}{l^2} - x^j x_j \right) \partial_t - \frac{4}{l} e^{-\frac{l}{2} t} x^i \partial_i,
  \end{split}
\end{equation}
where now $i,j=1,2,3$ and $x^i=(x,y,z)$. The $i,j$ indices are raised and lowered with the spatial part of the metric $g_{ij}=e^{l t} \delta_{ij}$. These vectors generate conserved charges only when $m=0$ (null geodesics).

Let us see how they can be distorted to generate conserved quantities in the massive case. First, let us construct the trivial Killing tensor $K=\partial_i \otimes \partial_i$ out of the sum of the tensor products of the Killing vectors that constitute the spatial translations. Its covariant components are
\begin{equation}
  K_{\mu\nu} = \begin{pmatrix}
                 0 & 0 & 0 & 0 \\
                 0 & e^{2l t} & 0 & 0 \\
                 0 & 0 & e^{2l t} & 0 \\
                 0 & 0 & 0 & e^{2l t}
               \end{pmatrix}
\end{equation}
and of course we have $\nabla_{(\kappa} K_{\mu\nu)}=0$. By using equation \eqref{geomups}, we find the following five vectors that satisfy it for appropriate functions $\Omega(x)$:
\begin{equation}\label{DSups}
  \begin{split}
  & \Upsilon_0 = \frac{e^{\frac{l}{2} t}}{\left(\frac{m^2}{\kappa} e^{l t}+1\right)^{\frac{1}{2}}} \partial_t, \quad \Upsilon_i = \frac{e^{\frac{l}{2} t}}{\left(\frac{m^2}{\kappa} e^{l t}+1\right)^{\frac{1}{2}}} x^i \partial_t - \frac{2}{l} e^{-\frac{l}{2} t} \left(\frac{m^2}{\kappa} e^{l t}+1\right)^{\frac{1}{2}} \partial_i \\
  & \Upsilon_4 = \frac{e^{-\frac{l}{2} t}}{\left(\frac{m^2}{\kappa} e^{l t}+1\right)^{\frac{1}{2}}} \left(\frac{4}{l^2} - x^j x_j \right) \partial_t - \frac{4}{l} e^{-\frac{l}{2} t} \left(\frac{m^2}{\kappa} e^{l t}+1\right)^{\frac{1}{2}}  x^i \partial_i.
  \end{split}
\end{equation}
Notice that by setting $m=0$ in \eqref{DSups} we obtain the proper conformal Killing vectors \eqref{DSCKV}. Due to theorem \ref{theo2} we expect the vectors $\tilde{\Upsilon}$, that emerge by substituting the constant ratio $\frac{m^2}{\kappa}$ in \eqref{DSups} with the expression
\begin{equation} \label{DScontodyn}
  \frac{m^2}{\kappa} = - \frac{g^{\mu\nu}p_\mu p_\nu}{K^{\alpha\beta}p_\alpha p_\beta}= \frac{p_\mu p^\mu}{p_x^2+p_y^2+p_z^2} = e^{-2 l t} \left(\frac{\dot{t}^2}{\dot{x}^2+\dot{y}^2+\dot{z}^2}-e^{l t}\right),
\end{equation}
are higher order symmetries of the Lagrangian. Truly, it can be easily verified that the resulting $\tilde{\Upsilon}(\x, \dot{\x}):=\Upsilon|_{\mathrm{eq} \; \eqref{DScontodyn}}$ satisfy $\mathrm{pr}^{(1)}\tilde{\Upsilon}(L)=0$ and form higher order Noether symmetries, whose charges are the $\tilde{I} = \tilde{\Upsilon}^\alpha p_\alpha$. As an example let us take the conserved charge corresponding to the
\begin{equation}\label{dsex1}
  \tilde{\Upsilon}_0 = \frac{e^{\frac{l t}{2}}}{\left(\frac{\dot{t}^2 e^{-l t}}{\dot{x}^2+\dot{y}^2+\dot{z}^2}\right)^{\frac{1}{2}}} \partial_t
\end{equation}
which reads
\begin{equation}\label{dsex1int}
  \tilde{I}_0= \tilde{\Upsilon}_0^\mu p_\mu = \tilde{\Upsilon}_0^\mu \frac{\partial L}{\partial \dot{\x}^\mu} = -\frac{e^{\frac{1}{2} l t} \dot{t}}{n \left(\frac{e^{-l t} \dot{t}^2}{\dot{x}^2+\dot{y}^2+\dot{z}^2}\right)^{\frac{1}{2}}}.
\end{equation}
It is straightforward to see that $\frac{d\tilde{I}_0 }{d\lambda}=0$ due to \eqref{DSeq1} and \eqref{DSeq2}. The corresponding on mass shell reduced expression of the charges $\tilde{I}$ are given by the $I=\Upsilon^\alpha p_\alpha$, which are conserved on account of $m^2 = \frac{1}{n} \left[\dot{t}- e^{l t}\left( \dot{x}^2+\dot{y}^2+\dot{z}^2 \right)\right]$ and $\kappa = \frac{e^{2l t}}{n} \left(  \dot{x}^2+\dot{y}^2+\dot{z}^2 \right) $.

We thus see that there exist geometries beyond pp-waves, where the conformal Killing vectors of the space admit mass dependent distortions. These generate additional conserved quantities when $m\neq 0$.

Up to now, we considered geodesics in (pseudo-)Riemannian geometry and saw how the symmetry breaking owed to the mass can lead to the appearance of new classes of symmetries. In the following sections we shall depart from the (pseudo-)Riemannian case, in order to introduce another symmetry breaking parameter. This will be realized through considering a more general, Finslerian, geometry and in particular that of a Bogoslovsky space-time, which involves a Lorentz violating parameter $b$.

\section{The Bogoslovsky-Finsler line-element}

We start this section with some generic information on Finsler geometry, so as to facilitate the following presentation. In Finsler geometry we consider a  general line-element of the form
\begin{equation} \label{Finel1}
  ds_F^2 = F(\x,d\x)^2,
\end{equation}
where $F(\x,d\x)$ is a function homogeneous of degree one in the $d\x^\mu$, that is, $F(\x,\sigma d\x) = \sigma  F(\x,d\x)$ for every $\sigma>0$. This is a generalization which contains the (pseudo-)Riemannian case, where the $F^2$ is simply quadratic in the $d\x^\mu$. A metric tensor can still be introduced as
\begin{equation} \label{Finmet}
  G_{\mu\nu}(\x,d\x) = - \frac{1}{2} \frac{\partial^2 F^2}{\partial(d\x^\mu)\partial (d\x^\nu)}
\end{equation}
and thus write
\begin{equation} \label{Finel2}
   ds_F^2 = G_{\mu\nu} (\x,d\x)  d\x^\mu d\x^\nu ,
\end{equation}
with the difference that the metric $G_{\mu\nu}$ now, in contrast to the Riemannian $g_{\mu\nu}$, carries a dependence on the differentials $d\x^\mu$. The equality of \eqref{Finel1} and \eqref{Finel2} is obtained with the use of Euler's theorem for homogeneous functions and from exploiting the fact that $F^2$ is a homogeneous function of degree two in the $d\x^\mu$. Given a Finsler function $F(\x,d\x)$ we may also express the line-element as
\begin{equation} \label{genFlineel}
  ds_F^2= F(\x,d\x)^2= \mathcal{F}(\x,d\x) g_{\mu\nu}(\x) d\x^\mu d\x^\nu
\end{equation}
where $\mathcal{F}(\x,d\x)$ is a function homogeneous of degree zero in $d\x$.

Bogoslovsky \cite{Bogo1,Bogo2} introduced a line-element of the form
\begin{equation} \label{bogoel}
  ds_F^2 =  \eta_{\mu\nu}d\x^\mu d\x^\nu \left[\frac{\left(\ell_\mu d\x^\mu\right)^{2}}{- \eta_{\mu\nu}d\x^\mu d\x^\nu}\right]^b,
\end{equation}
where $0<b<1$ is a dimensionless parameter, $\eta_{\mu\nu}=\mathrm{diag}(-1,1,1,1)$, $\nabla_\mu \ell^\nu=0$, $\eta_{\mu\nu}\ell^\mu \ell^\nu =0$ and $\ell^0>0$. This serves as a generalization of Special Relativity where the isotropy is broken in a preferred direction, which is set by the future directed null vector $\ell$. The symmetries of line-element \eqref{bogoel} have been studied in \cite{GiGoPo1} and are identified to form the eight dimensional group named $DISIM_b(2)$, which is a deformation of the $ISIM(2)$ group of Very Special Relativity \cite{VSL}. This lower symmetry count, in comparison to the ten dimensional Poincar\'e group of the quadratic line-element $ds^2=\eta_{\mu\nu} d\x^\mu d\x^\nu$, implies that we have another symmetry breaking effect owed to the anisotropy parameter $b$.

Bogoslovsky's theory has an obvious generalization to a curved space with the substitution $\eta_{\mu\nu}\mapsto g_{\mu\nu}$ \cite{Bogo3,Stavrinos}. In the case of a pp-wave space-time we can set as the preferred direction the covariantly constant null vector $\ell=\partial_v$. Then, the line-element of the Finslerian extension is written as
\begin{align} \label{bogoel2}
  ds_F^2  = g_{\mu\nu}d\x^\mu d\x^\nu \left[\frac{K_{\alpha\beta}d\x^\alpha d\x^\beta}{- g_{\mu\nu}d\x^\mu d\x^\nu}\right]^b ,
\end{align}
where, we made the use of the $K_{\alpha\beta}d\x^\alpha d\x^\beta=\left(\ell_\mu d\x^\mu\right)^{2}=du^2$ of the pp-wave case - remember that $K_{\mu\nu}=\ell_\mu \ell_\nu$.

For line-elements of the form of \eqref{bogoel2}, appearing in this Finslerian version of pp-waves, there exists an interesting theorem owed to Roxburgh and proven in \cite{Rox}. In brief it states that, if in the Finslerian line-element \eqref{genFlineel} the function $\mathcal{F}(\x,d\x)$ is such, so that
\begin{equation}
   \mathcal{F}(\x,d\x) = \mathcal{F} \left(\frac{\left(K_{\mu_1....\mu_k}(x) dx^{\mu_1}...dx^{\mu_k}\right)^{\frac{2}{k}}}{g_{\mu\nu}(\x) d\x^\mu d\x^\nu} \right) ,
\end{equation}
with $K_{\mu_1....\mu_k}(\x)$ a tensor of rank $k$, which is covariantly constant ($\nabla_\kappa K_{\mu\nu}=0$) with respect to the connection associated with $g_{\mu\nu}$, then the geodesics of $ds_F^2$ are identical to those produced by the Riemannian metric $g_{\mu\nu}$ and the typical quadratic line-element $ds^2=g_{\mu\nu}d\x^\mu d\x^\nu$.

Obviously, the pp-wave case falls into this category since $\ell$ is covariantly constant and hence so is $K_{\mu\nu}$. Thus, Finslerian pp-waves of this type basically produce the same geodesics as the Riemannian case. However, the physically related parameters are indeed affected by the presence of $b\neq 0$. What is more, we are going to see that interesting changes take place in what regards the symmetry structure of the system and - what the mass does when breaking the conformal Killing symmetries in the Riemannian case - now the parameter $b$ also does it to certain isometries of the base metric $g_{\mu\nu}$.

\section{Geodesics and a new conservation law}

The geodesic Lagrangian for the Finslerian line-element $ds_F^2$ of \eqref{bogoel} is given by
\begin{equation} \label{Lag1}
  L_F = -m\sqrt{-F^2} ,
\end{equation}
or equivalently, in the einbein formalism, by
\begin{equation}
  L = \frac{1}{2n} G_{\mu\nu}(\x,\dot{\x}) \dot{\x}^\mu \dot{\x}^\nu - n\frac{m^2}{2} ,
\end{equation}
which is of the same form as \eqref{primLag} with the difference that instead of $g_{\mu\nu}$, it now involves the Finsler metric $G_{\mu\nu}(\x,\dot{\x})$ defined in \eqref{Finmet}.

The (geodesic) Euler-Lagrange equations for the degrees of freedom $\x^\mu$ are equivalent to \cite{Baobook}
\begin{equation} \label{FinEL}
  \ddot{\x}^\mu + \gamma^\mu_{\kappa\lambda}\dot{\x}^\kappa \dot{\x}^\lambda =\x^\mu \frac{d}{d\lambda}\left(\ln n\right),
\end{equation}
where $\gamma^\mu_{\kappa\lambda} = \frac{1}{2} G^{\mu\sigma} \left( \frac{\partial G_{\sigma\lambda}}{\partial \x^\kappa} + \frac{\partial G_{\kappa\sigma}}{\partial \x^\lambda} -  \frac{\partial G_{\kappa\lambda}}{\partial \x^\sigma}\right)$ are the Christoffel symbols with respect to the metric $G_{\mu\nu}$. The Euler-Lagrange equation for the einbein field, $n$, yields the constraint
\begin{equation}
 \frac{1}{n^2}  G_{\mu\nu}(\x,\dot{\x}) \dot{x}^\mu \dot{x}^\nu + m^2 =0.
\end{equation}
The momenta are given by
\begin{equation} \label{Finmom}
  p_\kappa = \frac{\partial L_n}{\partial \dot{\x}^\mu} = \frac{1}{n} G_{\mu\kappa} \dot{\x}^\mu ,
\end{equation}
which looks similar to the relation of the Riemannian case, but this time the right hand side is not linear in the velocities. We note that the extra term that would appear, due to the dependence of $G_{\mu\nu}$ on the velocities, is identically zero by virtue of the definition \eqref{Finmet} and the fact that the Finsler metric, $G_{\mu\nu}$, is a homogeneous function of degree zero in the velocities; this term would be proportional to
\begin{equation}
  \frac{\partial G_{\mu\nu}}{\partial \dot{\x}^\kappa} \dot{\x}^\mu \dot{\x}^\nu = -\frac{1}{2}\frac{\partial^3 F^2}{\partial \dot{\x}^\mu\partial \dot{\x}^\nu\partial \dot{\x}^\kappa}\dot{\x}^\mu \dot{\x}^\nu = \frac{\partial G_{\mu\kappa}}{\partial \dot{\x}^\nu}\dot{\x}^\mu \dot{\x}^\nu =0 .
\end{equation}
The last equality holds due to Euler's theorem on homogeneous functions, which in this case implies $\frac{\partial G_{\mu\kappa}}{\partial \dot{\x}^\nu}\dot{\x}^\nu=0$ (see \eqref{Eultheo} for $k=0$).

We may thus once more write the Hamiltonian constraint as
\begin{equation} \label{FinHamcon}
  \mathcal{H} = G^{\mu\nu}(\x,p)p_\mu p_\nu + m^2 \approx 0,
\end{equation}
assuming that we have managed to invert relations \eqref{Finmom} and thus express the velocities with respect to the momenta. If we want to look for a linear in the momenta conserved quantity of the form $I=\Upsilon^\mu p_\mu$, we need to demand $\{I,\mathcal{H}\} \approx 0$. We use the weak equality here because, due to the Hamiltonian constraint \eqref{FinHamcon}, which is bound to be zero, it is enough that the Poisson bracket produces something multiple of the constraint. The Poisson bracket is calculated to be
\begin{equation} \label{PBraG}
  \begin{split}
  \{I,\mathcal{H}\} & = - \left( \Upsilon^\sigma \frac{\partial G^{\mu\nu}}{\partial \x^\sigma} + G^{\mu\sigma} \frac{\partial \Upsilon^\nu}{\partial \x^\sigma} + G^{\sigma\nu} \frac{\partial \Upsilon^\mu}{\partial \x^\sigma}\right) p_\mu p_\nu \\
   & = \frac{1}{n^2}\left( \Upsilon^\sigma \frac{\partial G_{\mu\nu}}{\partial \x^\sigma} - G_{\mu\sigma} \frac{\partial \Upsilon^\nu}{\partial \x^\sigma} - G_{\sigma\nu} \frac{\partial \Upsilon^\mu}{\partial \x^\sigma}\right) \dot{\x}^\mu \dot{\x}^\nu ,
  \end{split}
\end{equation}
where, in the last equality, we made the transition to velocity phase space coordinates by utilizing \eqref{Finmom}. In the parenthesis we recognize what would be the Lie derivative of $G_{\mu\nu}$ if the latter had no dependence in the velocities. Once more, an additional appearing term of the form $\frac{\partial \Upsilon^\kappa}{\partial \x^\sigma}\frac{\partial G^{\mu\nu}}{\partial p_\sigma}p_\mu p_\nu p_\kappa$ has been set to zero because
\begin{equation}
   \frac{\partial G^{\mu\nu}}{\partial p_\sigma}p_\mu = \frac{\partial G^{\mu\nu}}{\partial \dot{\x}^\kappa} \frac{\partial \dot{\x}^\kappa}{\partial p_\sigma} p_\mu = - G^{\lambda\nu} G^{\tau \mu} \frac{\partial G_{\lambda\tau}}{\partial \dot{\x}^\kappa} \frac{\partial \dot{\x}^\kappa}{\partial p_\sigma} G_{\rho\mu} \dot{\x}^\rho = - G^{\lambda\nu} \frac{\partial \dot{\x}^\kappa}{\partial p_\sigma} \frac{\partial G_{\lambda\rho}}{\partial \dot{\x}^\kappa}\dot{\x}^\rho=0 .
\end{equation}
The last equality holds again due to Euler's theorem.

For the line-element of \eqref{bogoel2} the corresponding Finsler metric $G_{\mu\nu}$ reads
\begin{equation} \label{Finmetsp}
   \begin{split}
     G_{\mu\nu}(\x,\dot{\x}) = & 2 b(1-b) \left[g_{\sigma \mu} g_{\tau \nu} \frac{\mathcal{K}^b}{\mathcal{G}^{1+b}} + \left( g_{\sigma\mu} K_{\tau\nu} + g_{\sigma\nu} K_{\tau\mu} \right) \frac{\mathcal{K}^{b-1}}{\mathcal{G}^b} + K_{\sigma \mu} K_{\tau \nu} \frac{\mathcal{K}^{b-2}}{\mathcal{G}^{1-b}}  \right] \dot{\x}^\sigma \dot{\x}^\tau \\
     & (1-b) g_{\mu\nu} \frac{\mathcal{K}^b}{\mathcal{G}^b}  - b K_{\mu\nu} \frac{\mathcal{K}^{b-1}}{\mathcal{G}^{b-1}},
   \end{split}
\end{equation}
where, for abbreviation, we use $\mathcal{K} = K_{\mu\nu}\dot{\x}^\mu \dot{\x}^\nu$ and $\mathcal{G} = -g_{\mu\nu}\dot{\x}^\mu \dot{\x}^\nu$. When we insert the metric \eqref{Finmetsp} inside \eqref{PBraG} and consider the equation $\{I,\mathcal{H}\}\approx 0$, we arrive at
\begin{equation} \label{intermcond}
  \{I,\mathcal{H}\} = \frac{1}{n^2} \left[(1-b) \left(\mathcal{L}_\Upsilon g_{\mu\nu}\right) \dot{\x}^\mu \dot{\x}^\nu \frac{\mathcal{K}^b}{\mathcal{G}^b}  - b \left(\mathcal{L}_\Upsilon K_{\mu\nu} \right) \dot{\x}^\mu \dot{\x}^\nu \frac{\mathcal{K}^{b-1}}{\mathcal{G}^{b-1}} \right] \approx  0 ,
\end{equation}
where $\mathcal{L}_\Upsilon$ now stands for the Lie derivative with respect to the vector $\Upsilon$.

In the case where $K_{\mu\nu}$ is a covariantly constant Killing vector, which means that, according to Roxburg's theorem \cite{Rox}, the \eqref{FinEL} become the Riemannian \eqref{eulgenLx}, the $\mathcal{K}$ and $\mathcal{G}$ are constants of the motion. Let us set the on mass shell constant value of their ratio as $\frac{\mathcal{K}}{\mathcal{G}}= \frac{1}{M_b^2}$. Then, the condition \eqref{intermcond} becomes
\begin{equation}
   \{I,\mathcal{H}\} = \frac{1}{n^2}\mathcal{L}_\Upsilon \left( \frac{1-b}{M_b^{2b}} g_{\mu\nu} - \frac{b}{M_b^{2(b-1)}} K_{\mu\nu}\right) \dot{\x}^\mu \dot{\x}^\nu \approx  0.
\end{equation}
The equality to zero is sufficient to be satisfied on mass shell. By taking an example from the Riemannian case, we may relax it to formulate the following:
\begin{theorem}
  Consider the Bogoslovsky space described by the line-element \eqref{bogoel2}. If $K$ is a second rank covariantly constant Killing tensor of $g$, and there exists a vector $\Upsilon$ satisfying
  \begin{equation}\label{Fincond}
    \mathcal{L}_\Upsilon \left( g_{\mu\nu} - \frac{b}{1-b} M_b^{2} K_{\mu\nu}\right) = 2 \Omega(\x) \left( g_{\mu\nu} + M_b^{2} K_{\mu\nu} \right),
  \end{equation}
  where the constant $M_b$ of the geodesic motion is defined as $M_b^{-2} = \frac{K_{\mu\nu} \dot{\x}^\mu \dot{\x}^\nu}{-g_{\mu\nu}\dot{\x}^\mu \dot{\x}^\nu}$, then the $I=\Upsilon^\mu p_\mu$ is conserved along the geodesics.
\end{theorem}
The proof is straightforward. When we contract the right hand side of equation \eqref{Fincond} with the velocities we obtain something which is on mass shell zero
\begin{equation}
  \{I,\mathcal{H}\} \propto \frac{2\Omega}{n^2} \left( g_{\mu\nu} + M_b^{2} K_{\mu\nu} \right) \dot{\x}^\mu \dot{\x}^\nu =\frac{2\Omega}{n^2}\left( -M_b^{2} K_{\mu\nu} \dot{\x}^\mu \dot{\x}^\nu + M_b^{2} K_{\mu\nu} \dot{\x}^\mu \dot{\x}^\nu \right) =0
\end{equation}
by simply using the fact that
\begin{equation} \label{Mbdef}
  M_b^{-2} = \frac{K_{\mu\nu} \dot{\x}^\mu \dot{\x}^\nu}{-g_{\mu\nu}\dot{\x}^\mu \dot{\x}^\nu}.
\end{equation}
For the condition \eqref{Fincond} we have not made any assumption about the space-time e.g. if it is going to be a pp-wave or not. The only thing that is needed is for $K$ to be covariantly constant with respect to the Levi-Civita connection compatible with $g_{\mu\nu}$, i.e. $\nabla_\kappa K_{\mu\nu}=0$. We may also notice that on the left hand side of \eqref{Fincond} there appears a metric disformally related to the original $g_{\mu\nu}$
\begin{equation}\label{disfmet2}
  \hat{g}_{\mu\nu} =  g_{\mu\nu} - \frac{b}{1-b} M_b^{2} K_{\mu\nu} .
\end{equation}
The Killing vectors of this metric, satisfy \eqref{Fincond} for $\Omega=0$. In the next section we proceed to study what happens in the case where $g$ is a pp-wave metric.

\section{The distorted symmetry vector in Finslerian pp-waves and in flat space}

Lets consider the metric $g$ of a pp-wave space-time derived from \eqref{lineel} and take as $K_{\mu\nu}$ the ``square'' of the covariantly constant vector $\ell=\partial_v$. A study on the integrability of the corresponding geodesic equations based on conventional integrals of motion has been given previously in \cite{BFgeo}. Here, we concentrate on the quantities that condition \eqref{Fincond} generates. The Finslerian geodesic Lagrangian \eqref{Lag1}, is written as
\begin{equation} \label{Lag1pp}
  L_F = -m\sqrt{-F^2} = - m \dot{u}^b \left(-H(u,x,y)\dot{u}^2 - 2 \dot{u}\dot{v} - \delta_{ij}x^i x^j \right)^{\frac{1-b}{2}} .
\end{equation}
Once more, we use the coordinates $\x^\mu = (u,v,x^i)$, as we did in section \ref{secRpp}. In the case $b=0$, the Lagrangian $L_F$ reduces to the usual square root Lagrangian of the Riemannian geodesics. The dynamically equivalent Lagrangian in the einbein formalism is given by
\begin{equation} \label{Lag2}
  L = -\frac{1}{2 n} \dot{u}^{2b} \left(-H(u,x,y)\dot{u}^2 - 2 \dot{u}\dot{v} - \delta_{ij}x^i x^j \right)^{1-b} - n \frac{m^2}{2}
\end{equation}
and it reproduces a set of Euler-Lagrange equations equivalent to those of $L_F$. The constraint equation (the Euler-Lagrange for $n$) leads to
\begin{equation}\label{ruln}
   n = \pm \frac{\dot{u}^b}{m} \left(-H(u,x,y)\dot{u}^2 - 2 \dot{u}\dot{v} - \delta_{ij}x^i x^j \right)^{\frac{1-b}{2}}.
\end{equation}
Substitution of \eqref{ruln} in \eqref{Lag2} gives $L= \pm L_F$. From now on - and to be consistent with the sign conventions assumed - where ever $n$ is substituted from \eqref{ruln} the plus root is utilized, so that we obtain the correspondence $L=L_F$.

Since $\ell$ is covariantly constant, obviously the same holds for $K = \ell \otimes \ell$, which we used to write \eqref{Lag1}. Thus, Lagrangians \eqref{Lag1pp} and \eqref{Lag2} are bound to generate the same geodesic equations as the \eqref{LagppR} of the Riemannian case. It can be easily verified that this is the true. Consequently, we expect that the same number of conserved quantities must be admitted in both systems. However, we need to mention, that it will not necessarily be the same vectors that generate symmetries. This is because, Lagrangians \eqref{Lag1pp} and \eqref{Lag2} are distinct from their Riemannian counterparts (obtained when $b=0$); they have a different functional dependence on velocities and this changes the form of the generating vectors.

An obvious symmetry that remains the same is the one owed to the covariantly constant vector $\ell$, which tells us that the momentum in the $v$ direction is again conserved. Truly, it is easy to see that we have the conserved charge
\begin{equation} \label{Finppintpv}
  p_v=\frac{\partial L}{\partial \dot{v}} = (1-b) \frac{\dot{u}^{1+2 b}}{n} \left(-H(u,x,y)\dot{u}^2 - 2 \dot{u}\dot{v} - \delta_{ij}x^i x^j \right)^{-b} = \pi_v.
\end{equation}
Once more, we use the Greek letter $\pi_v$ to denote the on mass shell constant value of the momentum $p_v$. By combining \eqref{Finppintpv} with \eqref{ruln} and remembering Eq. \eqref{Mbdef}, it is easy to derive the relation
\begin{equation} \label{bmass}
  M_b^2 = \left[ \frac{(1-b)^2 m^2}{\pi_v^2} \right]^{\frac{1}{1+b}}
\end{equation}
among the constants of integration. The latter, gives us $M_b$ in terms of the mass $m$, the momentum $\pi_v$ and the Lorentz violating parameter $b$; obviously, when $b=0$, $M_0^2 = \frac{m^2}{\pi_v^2}$.

By solving equation \eqref{Fincond}, we now obtain the following vector
\begin{equation}\label{symvec}
  \Upsilon^\mu = Y^\mu + M_b^2 f_1^\mu + \frac{b}{1-b} M_b^2 f_2^\mu,
\end{equation}
where $Y$ is a conformal Killing vector of the pp-wave metric $g_{\mu\nu}$ (its components are given by Eqs. \eqref{zetackv}) and $f_1$, $f_2$ are the acquired distortions. The first, is exactly the same as the one derived in \eqref{coefef} with the identification $\frac{m^2}{\pi_v^2}\rightarrow M_b^2$, i.e.
\begin{subequations} \label{modf1}
\begin{align}
  f_1^u  = & 0,
 \\ \nonumber
  f_1^v  = & \frac{1}{2} u  \left(x^i a_i'(u) - a' (u)+2 \bar{b}(u)-2 \mu v \right) + \frac{1}{2} x^i a_i (u) + \frac{\mu}{4} \delta_{ij} x^i x^j
 \\ \label{modf1v}
 & + \frac{a(u)}{2}  - M_b^2 \frac{\mu}{4} u^2\,,
  \\
f_1^{i} = & -\frac{1}{2}u \left(\mu\, x^i + a_i(u) \right).
\end{align}
\end{subequations}
The second distortion is new and contributes only along the $v$ direction
\begin{subequations} \label{modf2}
  \begin{align}
  & f_2^u =0 ,\quad f_2^i =0 \\ \label{modf2v}
  & f_2^v = \frac{\mu}{2} \delta_{ij}x^i x^j + a_i(u) x^i + a(u) .
  \end{align}
\end{subequations}
An alternative way to write the expression for $\Upsilon$ using only the $Y$ vector components is
\begin{equation} \label{symvec2}
  \Upsilon^\mu = Y^\mu + \sum_{n=1}^2 M_b^{2n} \frac{u^n}{2^{2n-1}}\frac{\partial^n}{\partial v^n} Y^\mu + \frac{b+1}{2 (1-b)} M_b^2 \; \delta^\mu_{\;v} Y^u .
\end{equation}
It can be checked that the quantity $I=\Upsilon^\mu \frac{\partial L}{\partial \dot{x}^\mu}$, made up from the vector \eqref{symvec}, is a constant of the motion, by virtue of the already known conserved charge \eqref{Finppintpv} and the constraint equation, which yields \eqref{ruln}. The corresponding function $\Omega(\x)$ for which $\Upsilon$ satisfies \eqref{Fincond} is given by
\begin{equation}
  \Omega(\x) = x^i a_i'(u) + \bar{b}(u) - \mu \left( v + \frac{M_b^2}{2} u \right) .
\end{equation}

A first observation regarding the vector \eqref{symvec} is that the Lorentz violating parameter $b$ introduces an additional distortion in terms of the vector $f_2$. In what follows we are going to study more in detail this distortion and its nature. Apart from the pp-wave case, we will also comment separately on what happens in the flat case, $H(u,x,y)=0$. The corresponding higher order symmetries of the flat case have been studied separately in \cite{Dimletter}.

Another interesting point is that, once more, if we take the basic vector \eqref{symvec} and substitute, in place of $M_b^2$, the ratio involving velocities, i.e. \eqref{Mbdef}. Then, exactly as it happened in the Riemannian case, the induced vector $\tilde{\Upsilon}:=\Upsilon|_{M_b^2=\frac{\mathcal{K}}{\mathcal{G}}}$ forms a higher order symmetry vector satisfying the infinitesimal criterion of invariance with $\mathrm{pr}^{(1)} \tilde{\Upsilon}(L)=0$. Hence, we again have a higher order Noether symmetry, whose on mass shell reduced expression yields the distorted space-time vector \eqref{symvec}.

\subsection{The non-flat case, $H(u,x,y)\neq 0$}

In the non flat case, where the metric describes a pp-wave space-time, the most general expression of a Killing vector, $\mathcal{L}_\xi g_{\mu\nu}=0$, is\footnote{We use the $\xi$ here to denote the subset of the $Y$ consisting only of the pure Killing vectors of the metric, i.e. those $Y$ corresponding to a conformal factor of $\omega(\x)=0$.} \cite{Maartens}
\begin{equation} \label{Kilgen}
  \xi = \left(\alpha u + \beta \right) \partial_u +  \left(\sigma -\alpha v - c_i'(u) x^i \right) \partial_v + \left(\gamma \epsilon_{ij}x^i + c_i(u) \right) \partial_j,
\end{equation}
whose components are obtained from \eqref{zetackv} when setting the functions, $a_i(u)$, $\bar{b}(u)$ and the parameter $\mu$ appearing in the conformal factor \eqref{omega} equal to zero and by introducing
\begin{equation} \label{nfkc}
  a(u) = \alpha u + \beta, \quad c(u) = \gamma \quad  \text{and} \quad  M(u,x,y) = \sigma - c_i'(u) x^i.
\end{equation}
Of course $H(u,x,y)$ has to also satisfy a certain partial differential equation, which is obtained from \eqref{rulM1} with the above substitutions together with $a_i(u)=\bar{b}(u)=\mu=0$
\begin{equation}
   \frac{1}{2} \left(\alpha u+\beta\right) \partial_u H + \frac{1}{2}\left(c_i(u) - \epsilon_{ij} x^j \right) \partial_i H + \alpha H - c_i''(u) x^i =0 .
\end{equation}
The constant parameters $\alpha, \beta, \gamma$ and $\sigma$ characterize the corresponding mono-parametric groups of motion; of them, the Killing vector owed to the parameter $\sigma$, i.e. $\xi_v= \ell=\partial_v$, is present in all pp-wave space-times. In total, a non-flat pp-wave can admit at most seven Killing vectors \cite{Maartens}.

As we observe from \eqref{modf1} and \eqref{modf2} the modification owed to the presence of Killing fields has just to do with the function $a(u)$, since it is the only one appearing in the $f_1^\mu$ and $f_2^\mu$ components, which at the same time does not belong to the conformal factor $\omega$ of \eqref{omega}. From the linear expression of $a(u)$ in \eqref{nfkc} we can also notice that no modification owed to the $f_1^\mu$ can affect a Killing vector. The $\alpha u$ part is automatically cancelled in the component $f_1^v$ by the combination $\frac{1}{2}\left(a-u a'\right)=\frac{\beta}{2}$; as it can be seen from \eqref{modf1v}. The remaining $\beta$ constant is nothing but a contribution which can be subtracted by a constant multiple of the already known Killing field $\xi_v=\ell=\partial_v$ and thus the $f_1^\mu$ can be cleared out from all modifications involving Killing fields.

On the contrary, the $f_2^\mu$ modification is bound to contain parameter $\alpha$ (when the appropriate Killing field exists), see the $f_2^v$ in \eqref{modf2v}. The parameter $\beta$ can still be removed by subtracting a multiple of the existing symmetry $\ell$. As a result, whenever $g_{\mu\nu}$ is such that the vector associated with $\alpha$,
\begin{equation}
  \xi_0 = u \partial_u - v \partial_v,
\end{equation}
is Killing, i.e. $\mathcal{L}_{\xi_0} g_{\mu\nu}=0$, then, $\xi_0$ is broken as a symmetry for the Finslerian space-time with $b\neq0$. However, with the appropriate distortion owed to $f_2$ we may write the
\begin{equation} \label{upsnfk}
  \Upsilon_0= u \partial_u + \left( \frac{b}{1-b} M_b^2 u- v \right)\partial_v,
\end{equation}
which generates a conserved charge $I=\Upsilon_0^\mu p_\mu$ in place of the broken symmetry. It is easy to check that, the $\Upsilon_0$ is a Killing vector of the disformally related metric \eqref{disfmet2}, i.e. $\mathcal{L}_{\Upsilon_0} \hat{g}_{\mu\nu}=0$. In other words, the modification that restores the broken Killing symmetry, satisfies \eqref{Fincond} for $\Omega =0$. We see thus that, in contrast to the $b=0$ case, when $b\neq 0$, it is possible to have a modification over a Killing vector field instead of just proper CKVs. The introduction of $b$ can truly break isometries and one needs to add certain modifications, which on the mass shell lead to conserved quantities.

From the Riemannian case, we remember that the distorted conformal Killing vectors - being the reduced form of some formal symmetries - do not necessarily close an algebra. Here, we shall see that the $b$-dependent modification acquired in $\Upsilon_0$ does not alter the algebra status with the rest of the symmetry vectors. To demonstrate this, let us break down the $\xi$ of \eqref{Kilgen} with respect to the rest of the parameters. Then, we have the following possible Killing vectors: $\xi_u=\partial_u$, $\xi_{xy} = -y \partial_x + x \partial_y$ and of course the always present $\xi_v=\ell=\partial_v$. In addition to the above we may also have vectors of the form
\begin{equation}
  \xi_{c_i} =  - c_i'(u) x^i  \partial_v  + c_i(u) \partial_j .
\end{equation}
It is clear from the above expressions that the only commutator relation that is altered by the modification term appearing in \eqref{upsnfk} is the
\begin{equation}
  [\Upsilon_0, \xi_u] = - \xi_u + \frac{b}{1-b} M_b^2 \xi_v .
\end{equation}
Even though this commutator brings about no problem in the closing of an algebra, in reality such a situation cannot arise for a non flat space-time, because $\xi_u$ and the non-modified vector $\xi_0 = u \partial_u - v\partial_v$ cannot be both Killing vectors for a non-flat metric of the form of $g_{\mu\nu}$ (application of both $\xi_u$ and $\xi_0$ leads to $H(u,x,y)=0$). Thus, the modified vector $\Upsilon_0$ is bound to close the same commutator relations as the original $\xi_0$ vector with the rest of the unbroken symmetries.

Finally, we can summarize that, in the pp-wave case, just one of the Killing vectors may acquire a distortion, the vector $\xi_0$. The distortion does not affect the property of closing an algebra with the rest of the Killing vectors of the original metric $g_{\mu\nu}$. Any other acquired distortions will be associated with the existing proper conformal Killing vectors.

\subsection{The flat case. ``Reinstating'' the Poincar\'e algebra.}

The flat space trivially satisfies the same relations as those of a pp-wave case by simply setting $H(u,x,y)=0$. The Bogoslovsky-Finsler line-element given by \eqref{bogoel} yields the space-time of Deformed Very Special Relativity, which, in light-cone coordinates, it is written
\begin{equation} \label{bogoelxp}
  ds_F^2 =  -\left(-2du dv -dx^2 - dy^2\right)^{1-b} (du)^{2b}.
\end{equation}
As we previously noticed, the geodesic motion is described either by \eqref{Lag1pp} or \eqref{Lag2} with the substitution $H(u,x,y)=0$. It is known that the space-time possesses an eight-dimensional symmetry group, the $DISIM_b(2)$, which was presented in \cite{GiGoPo1} and which is a deformation of the $ISIM(2)$ group of Very Special Relativity \cite{VSL}. The symmetry generators of $DISIM_b(2)$  are given by:
\begin{itemize}
  \begin{subequations} \label{disimb}
  \item  The translations
  \begin{equation}
    T_u = \partial_u, \quad T_v = \partial_v, \quad T_i = \partial_i
  \end{equation}
  \item the rotation
  \begin{equation}
    R = x \partial_y - y \partial_x,
  \end{equation}
  \item the (combination of boosts and rotations)
  \begin{equation}
    B_{ui}= u \partial_i - x^i \partial_v
  \end{equation}
  \end{subequations}
\end{itemize}
and a vector with an explicit dependence on $b$
\begin{equation} \label{Nb8}
\mathcal{N}_b = (b-1) u \partial_u +(1+b) v \partial_v + b x^i \partial_i .
\end{equation}
By looking at the \eqref{disimb}, we understand that the existence of the parameter $b$, has broken the three of the rest of the Poincar\'e symmetries, namely the vectors
\begin{subequations} \label{Poimiss}
\begin{align}
    B_0 &  = u \partial_u - v \partial_v \\
    B_{vi} & = -x^i \partial_u  + v \partial_x^j,
\end{align}
\end{subequations}
which now fail to produce conserved charges for the geodesic motion in the space characterized by \eqref{bogoelxp}.

In \cite{Dimletter}, it was shown that these symmetries are substituted by higher order symmetries, which are associated with the distorted vectors of the type that we study here. If we apply the condition \eqref{Fincond} for the metric $g_{\mu\nu}$ with $H(u,x,y)=0$ we derive - except from the known symmetries \eqref{disimb} - the following additional vectors
\begin{itemize}
\item The distorted Killing (for $b=0$) vectors
\begin{subequations} \label{distKil}
  \begin{align}
    & \Upsilon_0 = u \partial_u + \left(\frac{b}{1-b} M_b^2 u - v\right) \partial_v \\
    & \Upsilon_{vi} = v \partial_x^j -x^i \partial_u - \frac{b}{1-b} M_b^2 x^i  \partial_v  .
  \end{align}
  \end{subequations}
\item The distorted proper conformal Killing (when $b=0$, $m=0$) vectors
\begin{subequations} \label{distCKV}
\begin{align} \label{distCKV1}
   \Upsilon_D =& \left(M_b^2 u + 2 v\right) \partial_v + x^i \partial_i \\
   \Upsilon_K =& u^2 \partial_u + \frac{1}{2} \left(\frac{1+b}{1-b} M_b^2 u^2 - \delta_{ij}x^i x^j \right) \partial_v + u x^i \partial_i \\
   \Upsilon_{C_1} =& \frac{\delta_{ij}x^i x^j}{2}  \partial_u - \frac{1}{4} \left[ M_b^4 u^2 + M_b^2 \left( 4 u v - \frac{1-b}{1+b}\delta_{ij}x^i x^j \right)+ 4 v^2 \right] \partial_v   \nonumber \\
   & - \left(\frac{M_b^2}{2} u + v\right)x^i \partial_i \\
   \Upsilon_{C_2}^k =& u x^k \partial_u + \left(\frac{1}{1-b}M_b^2 u +v \right)x^k\partial_v - \frac{1}{2} \left(M_b^2 u^2 +2 u v +\delta_{ij}x^i x^j \right) \partial_k + x^k x^i \partial_i.
\end{align}
\end{subequations}
\end{itemize}
All of the above yield linear in the momenta conserved quantities - we shall see an example of this later. We notice that the first set \eqref{distKil}, consists of distortions of the Killing vectors \eqref{Poimiss} of $g_{\mu\nu}$. The $\Upsilon_0$ and $\Upsilon_{vi}$ are themselves Killing vectors, but of the disformally related metric $\hat{g}_{\mu\nu}$ of \eqref{disfmet2}, which in these coordinates reads
\begin{equation} \label{modmet0}
  \hat{g}_{\mu\nu} = \begin{pmatrix}
                 -\frac{b}{1-b}M_b^2 & 1 & 0 & 0 \\
                 1 & 0 & 0 & 0 \\
                 0 & 0 & 1 & 0 \\
                 0 & 0 & 0 & 1
               \end{pmatrix}.
\end{equation}
Thus, we can now understand how the condition \eqref{Fincond} works for the various vectors. The non-distorted symmetries \eqref{disimb} satisfy \eqref{Fincond} by yielding $\mathcal{L}_X g_{\mu\nu}=\mathcal{L}_X K_{\mu\nu}=0$, where $X$ is any vector of the \eqref{disimb}. The \eqref{distKil} satisfy \eqref{Fincond} by being Killing vectors of the disformally related metric \eqref{modmet0}, i.e. $\mathcal{L}_X \hat{g}_{\mu\nu}=0$, where $X$ is now any of the \eqref{distKil}. Finally, the \eqref{distCKV} are solutions of \eqref{Fincond} for appropriate non-zero $\Omega(\x)$.

An interesting additional observation is that the linear combination of the distorted homothecy $\Upsilon_D$ and Killing vector $\Upsilon_0$ gives rise to the symmetry vector \eqref{Nb8}, which together with the seven Killing vectors ($T_u,T_v,T_i,R,B_{ui}$) forms the eight dimensional algebra corresponding to the $DISIM_b(2)$ group of the symmetries we mentioned in the beginning of the section
\begin{equation}
  \mathcal{N}_b = b \Upsilon_D + (b-1) \Upsilon_0 .
\end{equation}

We already stated that the three \eqref{distKil} are Killing vectors of $\hat{g}_{\mu\nu}$, the same holds trivially also for the seven \eqref{disimb} since the action of their Lie derivative returns a zero for both $g_{\mu\nu}$ and $K_{\mu\nu}$. Hence, they are bound to close an algebra. The non-zero Lie brackets among these ten vectors are:
\begin{equation} \label{alg1}
  \begin{split}
    & [T_u,B_{ui}] = T_i, \quad [T_u,\Upsilon_0]=T_u + \frac{b}{1-b} M_b^2 T_v, \quad [T_v,\Upsilon_0]= -T_v \\
    & [T_v,\Upsilon_{vi}] = T_i, \quad [T_i,R]=\epsilon_{ij} T_j, \quad [T_i,B_{uj}]= -\delta_{ij} T_v \\
    & [T_i,B_{vj}] = - \delta_{ij}\left( T_u+ \frac{b}{1-b} M_b^2 T_v \right), \quad [R,B_{ui}] = \epsilon_{ji} B_{uj}\\
    & [R,\Upsilon_{vi}] = \epsilon_{ji} B_{vj}, \quad [B_{ui},\Upsilon_0] = - B_{ui}, \quad [B_{ui},B_{vj}] = \epsilon_{ji}R- \delta_{ij}\Upsilon_0 \\
    & [\Upsilon_0,\Upsilon_{vi}] = -\Upsilon_{vi}+\frac{b}{1-b} M_b^2 B_{ui}, \quad [\Upsilon_{vi},B_{vj}] = \epsilon_{ji} \frac{b}{1-b} M_b^2 R .
  \end{split}
\end{equation}
This is isomorphic to the Poincar\'e algebra. It can be easily noticed by simply observing that the disformally related metric $\hat{g}_{\mu\nu}$ is flat. Hence, its ten Killing vectors \eqref{disimb} and \eqref{distKil} just span the Poincar\'e algebra expressed in a different coordinate system. We need to be clear however, that the actual symmetry of the space with line-element \eqref{bogoelxp} is still the $DISIM_b(2)$ group. Strictly speaking, the \eqref{distKil} are not formal symmetries. As we already mentioned, such vectors (together with the \eqref{distCKV}) are the on mass shell reduced expressions of higher order symmetries, which happen upon the reduction to yield space-time vectors.

\subsubsection{Constants of the motion and comparison with the Minkowski case}

In order to make direct comparisons with the free relativistic particle in Minkowski space let us use the transformation
\begin{equation} \label{uvtotz}
  u = \frac{1}{\sqrt{2}} \left(z-t\right), \quad v = \frac{1}{\sqrt{2}} \left(t+z\right)
\end{equation}
and take some linear combinations of the vectors involved in the algebra \eqref{alg1}, so as to write them as
  \begin{equation} \label{genmink}
    \begin{split}
      \mathcal{T}_\mu & = \frac{\partial}{\partial x^\mu},  \\
      \mathcal{L}^{ij} & = x^j \frac{\partial}{\partial x^i} - x^i \frac{\partial}{\partial x^j} - \frac{b M_b^2}{2(1-b)} \left( \delta^j_z x^i -  \delta^i_z x^j\right) \left(\frac{\partial}{\partial t}+\frac{\partial}{\partial z} \right), \quad i,j=1,2,3 \\
      \mathcal{M}^j & = x^j \frac{\partial}{\partial t}+ t \frac{\partial}{\partial x^j} - \frac{b M_b^2}{2(1-b)} \left(x^j -  \delta^j_z t \right)\left(\frac{\partial}{\partial t}+\frac{\partial}{\partial z} \right) .
    \end{split}
  \end{equation}
Notice that here, for the Cartesian coordinates, we use the Latin indices to denote the spatial components. Thus, the $i,j$ in this section run from $1$ to $3$. We additionally observe that, in these coordinates, the only vectors that do not admit a modification based on $b$ are the translations and the rotation in the $x-y$ plane. Of course by linear combinations one can write as many ``unmodified by $b$" Killing vectors as in the previous section. However, we choose to use \eqref{genmink} as the basic vectors so that we have a direct comparison with what we know from the classical free relativistic particle problem, when $b=0$ is enforced. Thus, $\mathcal{L}^{ij}$ become the rotations and $\mathcal{M}^j$ the boosts when $b=0$.

Under the use of transformation \eqref{uvtotz} the Lagrangian \eqref{Lag2} becomes
\begin{equation} \label{flatexLag}
  L = - \frac{1}{2^{1+b} n } \left(\dot{z} - \dot{t}\right)^{2b} (-\eta_{\mu\nu}x^\mu x^\nu)^{1-b} - n \frac{m^2}{2}, \quad ,
\end{equation}
where $\eta_{\mu\nu}= \mathrm{diag}(-1,1,1,1)$. For $b=0$, we obviously recover the Lagrangian of a relativistic free particle in flat space. The solution to the Euler-Lagrange equations can be written as
\begin{subequations} \label{flatsol}
  \begin{align}
    & n(\lambda)= 2^{-\frac{b}{b+1}} (1-b)^{\frac{1-b}{1+b}} m^{-\frac{2 b}{b+1}} (p_0+p_z)^{\frac{2 b}{1+b}} \\
    & t(\lambda)= t_0- \frac{1}{2} \left[ \frac{p_x^2+p_y^2}{(p_0+p_z)^2} + 1 +  \frac{(1-b)^{\frac{2}{b+1}} m^{\frac{2}{b+1}}}{2^{\frac{b}{b+1}} (p_0+p_z)^{\frac{2}{b+1}}} \right]  (p_0+p_z)\lambda  \\
    & x(\lambda) =  p_x \lambda +x_0 \\
    & y(\lambda) =  p_y \lambda +y_0 \\
    & z(\lambda) =  z_0 - \frac{1}{2}  \left[\frac{p_x^2+p_y^2}{(p_0+p_z)^2} -1 +  \frac{(1-b)^{\frac{2}{b+1}} m^{\frac{2}{b+1}}}{2^{\frac{b}{b+1}}(p_0+p_z)^{\frac{2}{b+1}}} \right]  (p_0+p_z)\lambda,
  \end{align}
\end{subequations}
where we have substituted the $b$-dependent mass \eqref{bmass} as
\begin{equation} \label{rulM}
  M_b = \left[\frac{\sqrt{2}(1-b)m}{p_0+p_z} \right]^{\frac{1}{1+b}},
\end{equation}
since we have $p_v=\frac{p_0+p_z}{\sqrt{2}}$ from \eqref{uvtotz}. The $t_0$, $x^i_0$ together with all the $p_\mu$ are constants of integration. Since \eqref{flatexLag} produces equations equivalent to the Minkowski case, the solutions $t(\lambda)$, $x(\lambda)$, $y(\lambda)$ and $z(\lambda)$ are bound to be linear in $\lambda$ when the latter is the affine parameter, i.e. when $n(\lambda)=$const. However, the constants of integration are now associated in a different manner with respect to the physical observables, due to the presence of $b$. In \eqref{flatsol}, the constants of integration are arranged so that on mass sell we have
\begin{equation}
  \frac{\partial L}{\partial \dot{t}} =p_0, \quad \frac{\partial L}{\partial \dot{x}} =p_x , \quad \frac{\partial L}{\partial \dot{y}} =p_y, \quad \frac{\partial L}{\partial \dot{z}} =p_z .
\end{equation}
As we mentioned, the $p_\mu$ here are all constant. This is owed to the fact that the translations $\mathcal{T}_\mu$ are still symmetries of the problem. In this section we make no reference to phase-space formalism, so we do not distinct between the variables $p_\mu$ and their on mass shell constant values, we simply use $p_\mu$ to also denote the constants of integration. It can be seen that when $b=0$, and under the constraint
\begin{equation} \label{rulm}
  m^2 = p_0^2-p_x^2-p_y^2-p_z^2,
\end{equation}
the expressions \eqref{flatsol} reduce to the usual
\begin{equation}
  n(\lambda)=1, \quad t(\lambda)= t_0 - p_0 \lambda,  \quad x^i(\lambda)= x^i_0 + p_i \lambda.
\end{equation}
With the help of \eqref{flatsol} we may write the relation
\begin{equation}
  \eta^{\mu\nu} \frac{\partial L}{\partial \dot{x}^\mu}\frac{\partial L}{\partial \dot{x}^\nu} = -2^{-\frac{b}{b+1}} (1+b) (1-b)^{\frac{1-b}{b+1}} m^{\frac{2}{b+1}} (p_0 + p_z)^{\frac{2 b}{b+1}},
\end{equation}
which is exactly equivalent with the one given in \cite{GiGoPo1} for the Hamiltonian constraint (when considering the $p_\mu$ as phase-space variables). Obviously, upon setting $b=0$ we return as to the usual Hamiltonian constraint of (special) relativistic motion \eqref{rulm}.

It is easy to verify that the whole set of \eqref{genmink} produces conserved quantities. Whenever $b=0$ and the constraint among constants \eqref{rulm} is used, they become those generated by the Poincar\'e generators for the free relativistic particle. For example let us take the modified rotation around $x$ axis. According to  \eqref{genmink} the vector is written as
\begin{equation}
  \mathcal{L}^{yz} = - \frac{b M_b^2}{2(1-b)} y \frac{\partial}{\partial t} + z \frac{\partial}{\partial y} - \left[1 + \frac{b M_b^2}{2(1-b)} \right] y \frac{\partial}{\partial z} .
\end{equation}
We expect a constant of motion to be given by the quantity
\begin{equation}
  I_{yz} = (\mathcal{L}^{yz})^\mu \frac{\partial L}{\partial \dot{x}^\mu}
\end{equation}
which yields
\begin{equation}
  \begin{split}
    I_{yz} = & \frac{2^{-(1+b)}}{n\left(\dot{t}^2-\dot{x}^2-\dot{y}^2-\dot{z}^2\right)^{b}} \left(\dot{z}-\dot{t}\right)^{2 b-1}  \Bigg[2 (1-b) z \dot{y} \left(\dot{z}-\dot{t}\right) \\
    & - y \Big(b \left(M_b^2-2\right) \dot{t}^2-2 \left(b \left(M_b^2-1\right)+1\right) \dot{t} \dot{z}+b M_b^2 \dot{z}^2+2 \left( b \dot{x}^2+ b \dot{y}^2+\dot{z}^2\right)\Big) \Bigg] .
  \end{split}
\end{equation}
Direct use of \eqref{flatsol} demonstrates that $I_{yz}$ is indeed a constant of motion on mass shell, acquiring the value
\begin{equation}
  \begin{split}
    I_{yz} =  \frac{(p_0+p_z)^{-\frac{b+3}{b+1}}}{2^{\frac{2 b+1}{b+1}} } \Big[& 2^{\frac{b}{b+1}} (p_0+p_z)^{\frac{2}{b+1}} \left( \left(p_x^2+p_y^2-\left(p_0+p_z\right)^2\right)y_0 +2 p_y (p_0+p_z) z_0 \right) \\
    &+ (1-b)^{\frac{2}{b+1}} m^{\frac{2}{b+1}} (p_0+p_z)^2 y_0  \Big],
  \end{split}
\end{equation}
where \eqref{rulM} has been used so as the physical mass appears in the expression. Upon substitution of the latter from \eqref{rulm} and by setting $b=0$ the above relation becomes non-other but
\begin{equation}
  I_{yz}|_{b=0} =  z_0 p_y - y_0 p_z,
\end{equation}
which is the usual angular momentum in the $x$ direction.

The same is true of course for the conserved quantities constructed with the distorted conformal Killing vectors. For example, let us take the $\Upsilon_D$ of \eqref{distCKV1}, which in Cartesian coordinates, performing the change \eqref{uvtotz}, becomes
\begin{equation}
  \Upsilon_D = \left(t+z +\frac{1}{2} M_b^2 \left(z-t \right) \right) \left(\partial_t + \partial_z\right) + x \partial_x + y \partial y .
\end{equation}
With the use of \eqref{flatsol} we calculate the on mass shell constant value of the charge,
\begin{equation}
  I_{D} = \Upsilon_D^\mu p_\mu =  p_x x_0 +p_y y_0 + (p_0 +p_z) (t_0+z_0) + \frac{(1-b)^{\frac{2}{1+b}} m^{\frac{2}{1+b}}}{2^{\frac{b}{1+b}}} (p_0 +p_z)^{\frac{b-1}{1+b}} (z_0-t_0) ,
\end{equation}
where once more \eqref{rulM} has been used. By setting $b=0$ we obtain the mass distorted charge of the Minkowski case
\begin{equation}
  I_{D}|_{b=0} =  p_x x_0 +p_y y_0 + (p_0 +p_z) (t_0+z_0) + m^{2} \frac{z_0-t_0}{p_0+p_z}
\end{equation}
and further, for $m=0$ we obtain the conserved charge of the null geodesics generated by the corresponding pure conformal Killing vector.

The disformally related metric \eqref{modmet0} in the Cartesian coordinates becomes
\begin{equation} \label{modmet}
  \hat{g}_{\mu\nu} = \begin{pmatrix}
                 -\left(1 + \frac{b M_b^2}{2(1-b)} \right) & 0 & 0 & \frac{b M_b^2}{2(1-b)} \\
                 0 & 1 & 0 & 0 \\
                 0 & 0 & 1 & 0 \\
                 \frac{b M_b^2}{2(1-b)} & 0 & 0 & 1 - \frac{b M_b^2}{2(1-b)}
               \end{pmatrix},
\end{equation}
and of course the \eqref{genmink} are its Killing vectors. We can write \eqref{modmet} as
\begin{equation}\label{disfmetflat}
   \hat{g}_{\mu\nu} = \eta_{\mu\nu} - \frac{b M_b^2}{1-b} \partial_\mu \phi \partial_\nu \phi
\end{equation}
with the introduction of a scalar $\phi = \frac{1}{\sqrt{2}}\left(z-t \right)$, so as to be consistent with in the original definition by Bekenstein regarding disformal transformations. It can be seen that the light-like geodesics are not preserved when passing from $\eta_{\mu\nu}$ to $\hat{g}_{\mu\nu}$, for $b\neq 0$. However, the causal structure is not affected since, for any vector $A^\mu$ for which $\eta_{\mu\nu} A^\mu A^\nu <0$, we obtain $\hat{g}_{\mu\nu} A^\mu A^\nu <0$ as long as $0 <b <1$. Of course, $\hat{g}_{\mu\nu}$ and $\eta_{\mu\nu}$ both describe a flat space, we perform the aforementioned comparison of $\hat{g}_{\mu\nu} A^\mu A^\nu$ with $\eta_{\mu\nu} A^\mu A^\nu$ by considering the coordinate system fixed. In other words, we take $\hat{g}_{\mu\nu}$ and $\eta_{\mu\nu}$ to be different metrics written in the same coordinate system, not the same metric expressed in different coordinate systems. A comparison from this point of view is reasonable if we remember that the $b\neq0$ case actually describes motion in a Finslerian geometry; the $\hat{g}_{\mu\nu}$ we write here serves as a (pseudo-)Riemannian ``simulation'' of how the motion looks for a massive particle of mass $m$.

\begin{figure}[ht]
\hspace{2cm}\includegraphics[scale=0.6]{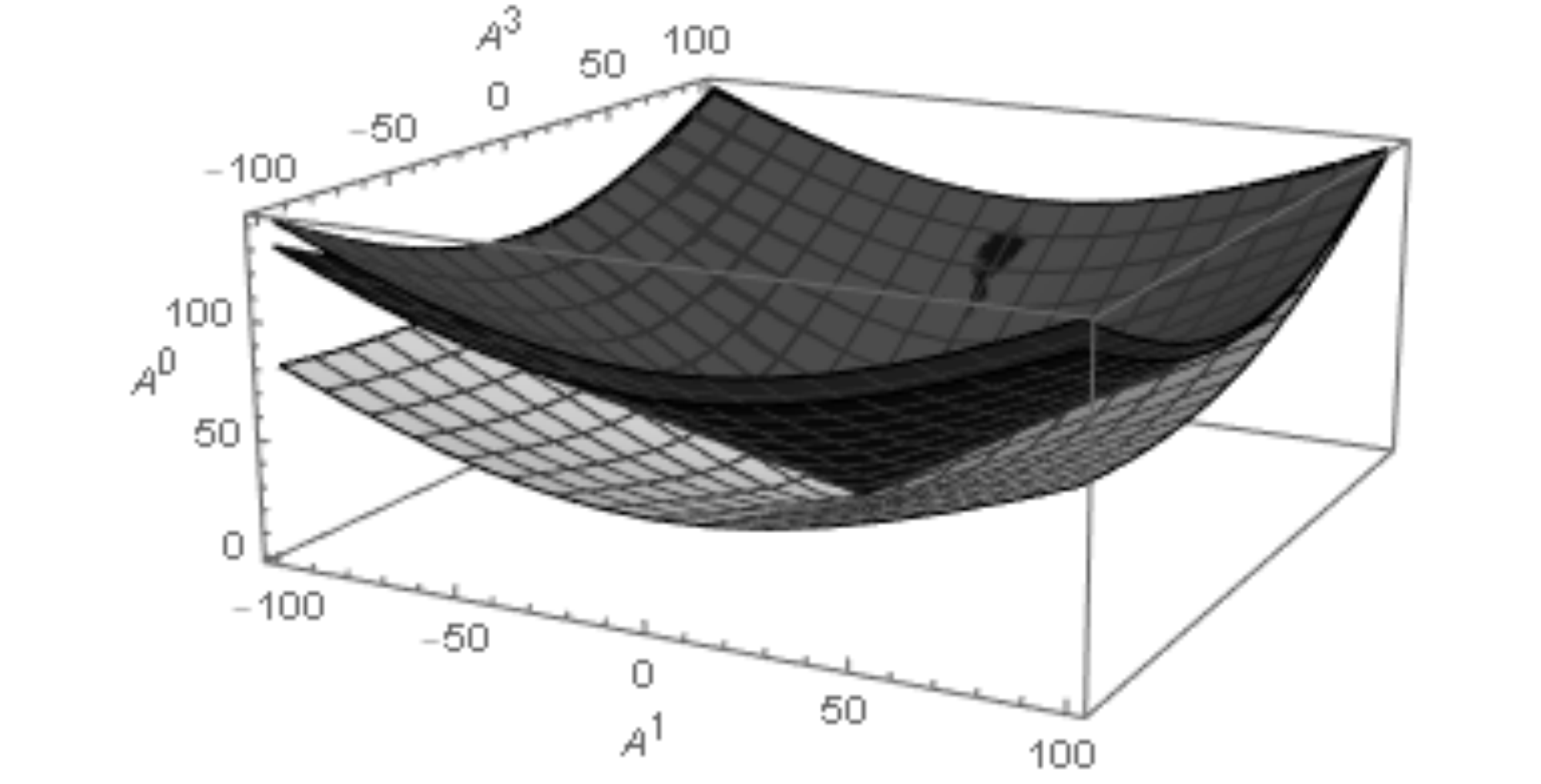}
\caption{Light cones in the $x-z$ plane for (starting from the upper surface) $b=0$, $b=\frac{1}{10}$ and $b=\frac{1}{2}$. In all cases we have considered $\frac{\sqrt{2} m}{p_0+p_z}=1$.\label{lcones}}
\end{figure}

In figure \ref{lcones} we draw the light cones\footnote{By using the term light cones here we do not imply the $m=0$ surfaces of the initial problem, but the geometric surfaces $\hat{g}_{\mu\nu} A^\mu A^\nu =0$ and how they differ from $\eta_{\mu\nu} A^\mu A^\nu =0$.} depending on the value of $b$ and given by the metric $\hat{g}_{\mu\nu}$, while keeping fixed the ratio where the physical mass $m$ is involved: $\frac{m}{\pi_v}=\frac{\sqrt{2} m}{p_0+p_z}=1$. The first layer from the top corresponds to $b=0$ where we have the typical Minkowski metric. The intermediate layer is for $b=\frac{1}{10}\Rightarrow M_b\simeq 0.995$ and the last corresponds to $b=\frac{1}{2}\Rightarrow M_b\simeq 0.909$. We observe the expected deviation in the $z$ direction from the isotropy and from the $b=0$ surface of Special Relativity as $b$ becomes larger. A physically reasonable value for the Lorentz violating parameter $b$ however, is way more minuscule $b<10^{-26}$ \cite{GiGoPo1}.

\section{Conclusion}

We examined how the elements of the conformal algebra of a given geometry may admit appropriate distortions, which lead to additional conserved quantities. We observed that these distortions are related to parameters that bring about an explicit symmetry breaking effect at the level of the geodesic equations. What is more, the resulting distorted vectors are generators of certain disformal transformations of the metric. We established a connection between these distorted conformal Killing vectors with higher order or hidden symmetries of the relative problem. The corresponding conserved Noether charges are in general rational functions of the momenta, which conveniently reduce to linear expressions on the mass shell.

We initiated our presentation by studying the geodesic motion in a (pseudo-)Riemannian space. The resulting distorted vectors in this case are owed to the mass, which breaks the proper conformal symmetries for non-null geodesics. The proper CKVs acquire mass dependent distortions in order to continue producing conserved quantities. We derived the geometric condition in order for such distortions to emerge. In short, the space-time needs to admit a second rank Killing tensor and additionally there has to exist a coordinate transformation, mapping the original metric to another, disformally related, which makes use of the same Killing tensor. Our basic example of a geometry satisfying the necessary conditions has been that of a generic pp-wave space-time. We wrote all the resulting distortions of the proper conformal Killing vectors and their connection to higher order symmetries. Apart from the pp-wave case however, we also presented an additional novel example in the form of the de Sitter solution of Einstein's equations with a cosmological constant. In the pp-wave case the distortion appeared in an additively manner, while in the de Sitter case it assumes a more complicated form.

The consideration of a Finslerian geometry in the form of the generalized Bogoslovsky line-element, revealed that the additional introduction of a (Lorentz) symmetry breaking parameter follows a similar pattern. This time, again for a pp-wave space-time, one of the Killing symmetries is lost due to the newly introduced parameter. The latter becomes involved in an appropriate distortion, which reveals a conserved quantity that takes the place of the one lost from the missing symmetry. For the rest of the proper conformal Killing vectors, similar distortions to those of the Riemannian case appear. We derive explicitly all the relative expressions and we notice that again the ensuing vectors are associated with higher order symmetries. We further investigated the general geometric conditions that are necessary so that such a type of symmetry emerges. Finally, we briefly mentioned how all this applies to the flat case, where three of the original Killing vectors acquire the necessary distortions. The distorted Killing vectors lead naturally to a disformally related metric by assuming the role of its isometries. We use this exact metric to compare the motion of the Finslerian case to the one taking place in the Minkowski space-time of Special Relativity.

By looking at the necessary condition that we derived for the Finslerian line-element, Eq. \eqref{Fincond}, we notice that, in contrast to the (pseudo-)Riemannian case, there exists an additional restriction; the condition requires the Killing tensor admitted by the base metric to be covariantly constant. This is easily satisfied in the pp-wave case, by simply utilizing the trivial tensor $K=\ell\otimes\ell$, constructed by the covariantly constant Killing vector $\ell$, which all pp-waves possess. However, the particular example of the de Sitter space, that we considered in the Riemannian case, does not apply here, since the relevant Killing tensor that we used there is not covariantly constant. This leaves an open question on whether we can incorporate other geometries in the Finslerian case, which can make use of \eqref{Fincond}.

It is particularly interesting how the pp-waves appear to be on the spot in what regards the emergence of this type of symmetries. They conveniently satisfy all the necessary conditions, both in the Riemannian case and in the Finslerian generalization. This adds up to the intriguing geometrical properties that these space-times possess and justifies their importance in physical theories. Further study is necessary however, in order to reveal other types of geometries where hidden symmetries can be reduced in such a way so as to be mapped to distortions of the conformal structure. The example of the de Sitter metric in the Riemannian case shows that this is in general possible and novel symmetries can be revealed for certain space-times.

From a pure mathematical perspective, the use of such hidden symmetries can be exploited to extend the known integrable classes of geodesic systems. The reduced linear integrals of the Lorentzian case, which are associated with the existence of proper conformal Killing vectors, imply that, for those space-times for which such quantities appear, the problem of the time-like geodesics has as many integrals of the motion as the null case. In \cite{Car3} it was demonstrated that the integrability of the geodesic system is closely related to that of the geodesic deviation equation. It is interesting to study if the hidden symmetries presented here can be ``inherited'' at the level of the geodesic deviation. Especially for the pp-wave case this could contribute to the further analytic study of the memory effect \cite{Gib1,Gib2,Shore,Chakra}.

In what regards the physical interpretation of each conserved charge, this highly depends on the exact space-time where the motion takes place. However, the fact that the mass of the particle appears explicitly in the expressions, allows for a direct comparison between the relative observables of the two cases (massive and massless). The same is true for the Finslerian case and the Lorentz violating parameter $b$. We saw in \eqref{distKil}, how the $b\neq0$ case affects the boosts and, of course, the corresponding integrals of the motion. We thus acquire a picture of how symmetry breaking parameters affect classical observables. In this respect it is intriguing how the explicit symmetry breaking, either because of the mass or due to some Lorentz violating parameter, leads to the appearance of hidden symmetries in the relevant theories. The latter seem to substitute the ones which were broken by the introduction of the corresponding parameter. This is a subject that certainly requires further attention and the study of additional examples.

Another interesting implication of the actual hidden symmetries is their possible realization at the quantum level. In several cases, hidden symmetries have been used in the context of a canonical quantization procedure: from the derivation of the hydrogen spectrum with the use of the Laplace-Runge-Lenz vector \cite{Pauli}, to modern problems in quantum cosmology \cite{Jala}. For quadratic Hamiltonians and higher order symmetries related to the existence of Killing tensors, there are various works, which explore a formal way of constructing the relative quantum operators under appropriate geometric conditions \cite{QM1,QM2,QM3}. It is true that there are cases where the quantum analogues of hidden symmetries - unlike their classical counterparts - do not commute with the Hamiltonian. This effect is referred in the literature as quantum anomaly \cite{Plylast}. In our case, we have to recognize an additional difficulty owed to the fact that the Noether charges are rational functions in the momenta. Thus, the construction of the corresponding quantum observables is far from trivial. Nevertheless, it is useful to further investigate if the on mass shell reduced linear expressions can be used in this respect, or if under appropriate (not affecting the quantum description) canonical transformations, one can obtain more manageable expressions.

\section*{Acknowledgements}
This work was supported by the Fundamental Research Funds for the Central Universities, Sichuan University Full-time Postdoctoral Research and Development Fund No. 2021SCU12117

\appendix

\section{Integrability conditions of $M(u,x,y)$} \label{app0}

The integrability conditions that need to be satisfied by the function $M(u,x,y)$, appearing in the components \eqref{zetackv} of a general conformal pp-wave Killing vector, are
\begin{subequations} \label{intM}
\begin{equation} \label{rulM1}
  \begin{split}
    \partial_u M(u,x,y) = & \left(\bar{b}(u) - a'(u) \right) H(u,x,y) - \frac{1}{2} \left( \frac{\mu}{2} \delta_{ij} x^i x^j + x^i a_i(u) + a(u)  \right) \partial_u H(u,x,y) \\
 & - \frac{1}{2} \left[ \left( \bar{b}(u) x^j + c(u) \epsilon_{ij} x^i + c_j(u) + \frac{1}{2} \gamma_{ijkl} a_i'(u) x^k x^l \right) \partial_j H(u,x,y) \right]
  \end{split}
\end{equation}
\begin{equation}
  \begin{split}
    \partial_i M(u,x,y) =&  - \left( \mu H(u,x,y)+ \bar{b}'(u) \right) x^i - a_i(u) H(u,x,y)+ c'(u) \epsilon_{ij} x^j + c_i'(u) \\
    & - \gamma_{ijkl} a_j''(u) x^k x^l .
  \end{split}
\end{equation}
\end{subequations}
In order to avoid confusion, we stress here, that one may note some differences over some constant factors, when comparing with the relative relations derived in the \cite{Maartens} paper. This is owed to the fact, that in the latter, the authors consider a slightly different line-element for the pp-wave space-time, than what we assume in \eqref{lineel}. Both \eqref{zetackv} and \eqref{intM}, which we use in this work, are compatible with \eqref{lineel} and differ slightly from the formalism encountered in \cite{Maartens}, due to the difference in the line-elements.

\section{Proof of theorem \ref{theo2}} \label{App1}

We assume that a vector of the form $\Upsilon=\Upsilon (x,\frac{m^2}{\kappa})$ solves \eqref{geomups}. Let us consider the vector $\tilde{\Upsilon}$ of \eqref{geomupshigh}, which for simplicity we will write in the form $\tilde{\Upsilon}=\tilde{\Upsilon}(\x,\frac{\mathcal{G}}{\mathcal{K}})$, understanding that we do not consider here $\mathcal{G}=-g_{\mu\nu} \dot{\x}^\mu \dot{\x}^\nu$ and $\mathcal{K}=K_{\mu\nu} \dot{\x}^\mu\dot{\x}^\nu$ as constants, but as quadratic functions in the velocities. The first prolongation of $\tilde{\Upsilon}$ is given by
\begin{equation} \label{apppro}
  \mathrm{pr}^{(1)} \tilde{\Upsilon} = \tilde{\Upsilon}^\kappa \frac{\partial}{\partial \x^\kappa} +  \dot{\tilde{\Upsilon}}^\kappa \frac{\partial}{\partial \dot{\x}^\kappa} , 
\end{equation}
where
\begin{equation} \label{appsub0}
  \dot{\tilde{\Upsilon}}^\kappa = \frac{\partial \tilde{\Upsilon}^\kappa}{\partial \x^\nu} \dot{\x}^\nu + \frac{\partial \tilde{\Upsilon}^\kappa}{\partial \dot{\x}^\nu} \ddot{\x}^\nu .
\end{equation}
We can split the partial derivative with respect to the positions $\x^\mu$ in two parts: one that differentiates the $\x$ dependence outside the $\frac{\mathcal{G}}{\mathcal{K}}$ ratio, i.e. $\frac{\partial \tilde{\Upsilon}^\mu}{\partial \x^\nu}\big|_{\left(\frac{\mathcal{G}}{\mathcal{K}}\right)}$, as if $\frac{\mathcal{G}}{\mathcal{K}}=$const., and another which contains separately the derivation of the $\frac{\mathcal{G}}{\mathcal{K}}$ part. So, we write
\begin{equation} \label{appsub1}
  \frac{\partial \tilde{\Upsilon}^\kappa}{\partial \x^\nu} = \frac{\partial \tilde{\Upsilon}^\kappa}{\partial \x^\nu}\bigg|_{\left(\frac{\mathcal{G}}{\mathcal{K}}\right)} -g_{\alpha\beta,\nu} \frac{\partial \tilde{\Upsilon}^\kappa}{\partial \mathcal{G}} \dot{\x}^\alpha \dot{\x}^\beta + K_{\alpha\beta,\nu} \frac{\partial \tilde{\Upsilon}^\kappa}{\partial \mathcal{K}} \dot{\x}^\alpha \dot{\x}^\beta.
\end{equation}
In addition, for the derivative of $\tilde{\Upsilon}^\kappa$ with respect to the velocities we have
\begin{equation} \label{appsub2}
  \frac{\partial \tilde{\Upsilon}^\kappa}{\partial \dot{\x}^\nu}  = - 2 g_{\alpha\nu} \frac{\partial \tilde{\Upsilon}^\kappa}{\partial \mathcal{G}} \dot{\x}^\alpha + 2 K_{\alpha\nu} \frac{\partial \tilde{\Upsilon}^\kappa}{\partial \mathcal{K}} \dot{\x}^\alpha  .
\end{equation}
By using \eqref{appsub1} and \eqref{appsub2} in \eqref{appsub0} we obtain
\begin{equation}
  \begin{split}
    \dot{\tilde{\Upsilon}}^\kappa = & \frac{\partial \tilde{\Upsilon}^\kappa}{\partial \x^\nu}\bigg|_{\left(\frac{\mathcal{G}}{\mathcal{K}}\right)} + \left( \nabla_{(\nu} K_{\alpha\beta)} \frac{\partial \tilde{\Upsilon}^\kappa}{\partial \mathcal{K}} -  \nabla_{(\nu} g_{\alpha\beta)} \frac{\partial \tilde{\Upsilon}^\kappa}{\partial \mathcal{G}} \right) \dot{\x}^\alpha \dot{\x}^\beta \dot{\x}^\nu \\
    & +2 \left( \mathcal{G}  \frac{\partial \tilde{\Upsilon}^\kappa}{\partial \mathcal{G}} + \mathcal{K} \frac{\partial \tilde{\Upsilon}^\kappa}{\partial \mathcal{K}} \right) \frac{d}{d\lambda} \left( \ln n \right),
  \end{split}
\end{equation}
where we have also used \eqref{eulgenLx} to substitute the accelerations $\ddot{\x}^\nu$ appearing in \eqref{appsub0}.

All but the first term in the above expression are zero: First of all, we have trivially $\nabla_{\nu} g_{\alpha\beta}=0$, where the $\nabla_\nu$ denotes the covariant derivative with respect to the Christoffel symbols. By our demand, we also have $\nabla_{(\nu} K_{\alpha\beta)}=0$, since $K$ is a Killing tensor. As for the relation, $\mathcal{G}  \frac{\partial \tilde{\Upsilon}^\kappa}{\partial \mathcal{G}} + \mathcal{K} \frac{\partial \tilde{\Upsilon}^\kappa}{\partial \mathcal{K}} =0$, it is zero due to the dependence of $Y^\kappa$ on the ratio $\frac{\mathcal{G}}{\mathcal{K}}$. Hence, the first prolonged vector \eqref{apppro} is written as
\begin{equation} \label{apppro2}
  \mathrm{pr}^{(1)} \tilde{\Upsilon} = \tilde{\Upsilon}^\kappa \frac{\partial}{\partial \x^\kappa} +   \frac{\partial \tilde{\Upsilon}^\kappa}{\partial \x^\nu}\bigg|_{\left(\frac{\mathcal{G}}{\mathcal{K}}\right)} \frac{\partial}{\partial \dot{\x}^\kappa} .
\end{equation}
The action of \eqref{apppro2} on the Lagrangian \eqref{primLag} yields
\begin{equation} \label{appproonL}
  \begin{split}
    \mathrm{pr}^{(1)}\tilde{\Upsilon}(L) = \frac{1}{2n} \left( \tilde{\Upsilon}^\kappa \frac{\partial}{\partial \x^\kappa} g_{\mu\nu} + g_{\mu\kappa} \frac{\partial \tilde{\Upsilon}^\kappa}{\partial \x^\nu}\bigg|_{\left(\frac{\mathcal{G}}{\mathcal{K}}\right)} + g_{\nu\kappa} \frac{\partial \tilde{\Upsilon}^\kappa}{\partial \x^\mu}\bigg|_{\left(\frac{\mathcal{G}}{\mathcal{K}}\right)} \right) \dot{\x}^\mu \dot{\x}^\nu .
  \end{split}
\end{equation}
What we see inside the parenthesis, is the left hand side of \eqref{geomups}, i.e. the $\Upsilon^\kappa \frac{\partial}{\partial \x^\kappa} g_{\mu\nu} + g_{\mu\kappa} \frac{\partial \Upsilon^\kappa}{\partial \x^\nu} + g_{\nu\kappa} \frac{\partial \Upsilon^\kappa}{\partial \x^\mu}$, since the remaining derivatives of $\tilde{\Upsilon}$ treat the ratio $\frac{\mathcal{G}}{\mathcal{K}}= \frac{-g_{\alpha\beta}\dot{\x}^\alpha \dot{x}^\beta}{K_{\mu\nu}\dot{\x}^\mu \dot{x}^\nu}=\frac{m^2}{\kappa}$ as a constant. As a result, with the use of equality \eqref{geomups}, we may write
\begin{equation}
  \mathrm{pr}^{(1)}\tilde{\Upsilon}(L) = \frac{\Omega(\x)}{n} \left( g_{\mu\nu} + \frac{\mathcal{G}}{\mathcal{K}} K_{\mu\nu}\right) \dot{\x}^\mu \dot{\x}^\nu = \frac{\Omega(\x)}{n} \left(- \mathcal{G}+ \frac{\mathcal{G}}{\mathcal{K}}\mathcal{K} \right)=0.
\end{equation}
Thus, the vector $\tilde{\Upsilon}(\x,\frac{\mathcal{G}}{\mathcal{K}})$ is a higher order symmetry of the geodesics, by virtue of \eqref{geomups} and because of $K$ being a Killing tensor. Its on mass shell reduced form is given by the vector $\Upsilon(\x,\frac{m^2}{\kappa})$.

The expression for the conserved charge \eqref{geomtIhigh}, generated by the symmetry $\tilde{\Upsilon}$, stems directly from Noether's theorem and in particular relation \eqref{genint} (our symmetry vector, $\tilde{\Upsilon}$, satisfies \eqref{generalsymcon} for $\Phi=$const. and has only components in the $\x$ directions, not in $\lambda$, so $\chi=0$).


\begin{thebibliography}{99}

\bibitem{Synge} J. L. Synge, MNRAS \textbf{131}, (1966) 463

\bibitem{Maeda} K. Hioki and K. Maeda, Phys. Rev. D \textbf{80}, (2009) 024042

\bibitem{Mann} S.-W. Wei, Y.-C. Zou, Y.-X. Liu and R. B. Mann, JCAP 1908, (2019) 030

\bibitem{Haitang} G. Guo, P. Wang, H. Wu and H. Yang, JHEP \textbf{2022}, (2022) 60

\bibitem{BHshadCh} Z. Zhang, H. Yan, M. Guo and B. Chen, arXiv preprint: 2206.04430 [gr-qc]

\bibitem{Gib1} P.-M. Zhang, C. Duval, G. W. Gibbons and P. A. Horvathy, Phys. Lett. B \textbf{772}, (2017) 743

\bibitem{Gib2} P.-M. Zhang, C. Duval, G. W. Gibbons and P. A. Horvathy, Phys. Rev. D \textbf{96}, (2017) 064013

\bibitem{Shore} G. M. Shore, JHEP \textbf{2018}, (2018) 133

\bibitem{Chakra} I. Chakraborty and S. Kar, Phys. Lett. B \textbf{808}, (2020) 135611

\bibitem{Katzin} G. H. Katzin, J. Math. Phys. \textbf{22}, (1981) 1878

\bibitem{Cav} G. Caviglia,  J. Math. Phys. \textbf{24}, (1983) 2065

\bibitem{Hoj} S. Hojman, L. Nu\~nez, A. Patiño and H. Rago, J. Math. Phys. \textbf{27}, (1986) 281

\bibitem{Rosquist} K. Rosquist, J. Math. Phys. \textbf{30}, (1989) 2319

\bibitem{Andr1} M. Tsamparlis and A. Paliathanasis, Gen. Rel. Grav. \textbf{42}, (2010) 2957

\bibitem{Andr2} M. Tsamparlis and A. Paliathanasis, Gen. Rel. Grav. \textbf{43}, (2011) 1861

\bibitem{Gomis} C. Batlle, J. Gomis, S. Ray and J. Zanelli,  Phys.Rev.D \textbf{99}, (2019) 064015

\bibitem{PetHar} P.-M. Zhang, M. Cariglia, M. Elbistan, G. W. Gibbons and P. A. Horvathy, Phys. Lett. B \textbf{792}, (2019) 324

\bibitem{Andr3} A. Paliathanasis, Symmetry \textbf{13}, (2021) 1018

\bibitem{Carter} B. Carter, Commun. Math. Phys. \textbf{10}, (1968) 280

\bibitem{Penrose} M. Walker and R. Penrose, Commun. Math. Phys. \textbf{18}, (1970) 265

\bibitem{Woodhouse} N. M. J. Woodhouse, Commun. Math. Phys. \textbf{44}, (1975) 9

\bibitem{Frolov} V. P. Frolov, P. Krtou\v{s} and D. Kubiz\v{n}\'ak, Living. Rev. Relativ. \textbf{20}, (2017) 6

\bibitem{Goldstein} H. Goldstein, C. Poole and J. Safko, ``\textit{Classical Mechanics}'', 3rd ed., Addison Wesley, San Francisco, Boston, New York, pp. 102 (2000)

\bibitem{Fradkin} D. M. Fradkin, Am. J. Phys. \textbf{33}, (1965) 207

\bibitem{AA1} A. A. Andrianov, M. V. Ioffe and D. N. Nishnianidze, Phys. Lett. A \textbf{201} (1995) 103

\bibitem{AA2} A. A. Andrianov, F. Cannata, M. V. Ioffe and D. N. Nishnianidze, J. Phys. A: Math. Gen. \textbf{30}, (1997) 5037

\bibitem{M1} M. Plyushchay, Int. J. Mod. Phys. A \textbf{15}, (2000) 3679

\bibitem{M2} M. S. Plyushchay, Phys. Lett. B \textbf{485}, (2000) 187

\bibitem{M3} C. Leiva and M. S. Plyushchay, Phys. Lett. B, \textbf{582}, (2004) 135

\bibitem{M4} M. S. Plyushchay, ``\textit{Nonlinear Supersymmetry as a Hidden Symmetry}'' in ``\textit{Integrability, Supersymmetry and Coherent States}'', S. Kuru, J. Negro and L. M. Nieto (Eds.), Springer Nature Switzerland, pp. 163 (2019)

\bibitem{Kalotas} T. M. Kalotas and B. G. Wybourne, J. Phys. A: Math. Gen. \textbf{15}, (1982) 2077

\bibitem{Carhid} M. Cariglia, Rev. Mod. Phys. \textbf{86}, (2014) 1283

\bibitem{Katzin2} G. H. Katzin and J. Levine, J. Math. Phys. \textbf{15}, (1974) 1460

\bibitem{PWHiggs} P. W. Higgs, J. Phys. A: Math. Gen. \textbf{12}, (1979) 309

\bibitem{Pal} G. Papagiannopoulos, J. D. Barrow, S. Basilakos, A. Giacomini and A. Paliathanasis, Phys. Rev. D \textbf{95}, (2017) 024021

\bibitem{Ply} L. Inzunza, M. S. Plyushchay and A. Wipf, JHEP \textbf{2020}, (2020) 028

\bibitem{Ply2} L. Inzunza and M. S. Plyushchay, JHEP \textbf{2021}, (2021) 165

\bibitem{Ply3} L. Inzunza and M. S. Plyushchay, JHEP \textbf{2022}, (2022) 179

\bibitem{Tsamp1} M. Tsamparlis and A. Mitsopoulos, J. Math. Phys. \textbf{61}, (2020) 072703

\bibitem{Tsamp2} A. Mitsopoulos and M. Tsamparlis, J. Geom. Phys. \textbf{170}, (2021) 104383

\bibitem{Liv3} J. B. Achour, E. R. Livine, S. Mukohyama, and J.-P. Uzan, JHEP \textbf{2022}, (2022) 112

\bibitem{dimgeo} N. Dimakis, Petros A. Terzis and T. Christodoulakis Phys. Rev. D \textbf{99}, (2019) 104061

\bibitem{pp1} K. Vesel\'y and J. Podolsk\'y, Phys. Rev. D \textbf{58}, (1998) 081501

\bibitem{pp2} A. Lecke, R. Steinbauer and R. \v{S}varc, Gen. Rel. Grav. \textbf{46}, (2014) 1648

\bibitem{pp3} U. Camci and A. Yildirim, Phys. Scripta \textbf{89}, (2014) 084003

\bibitem{pp4} J. W. Maluf, J. F. da Rocha-Neto, S. C. Ulhoa and F. L. Carneiro, JCAP 03, (2019) 028

\bibitem{pp5} P.-M. Zhang, M. Cariglia, M. Elbistan and P. A. Horvathy, J. Math. Phys. \textbf{61}, (2020) 022502

\bibitem{pp6} K. Vesel\'y and J. Podolsk\'y, Phys. Lett. A \textbf{271}, (2000) 368

\bibitem{pp7} M. Elbistan, P.-M. Zhang, G. W. Gibbons and P. A. Horvathy, JCAP 01, (2021) 052

\bibitem{pp8} M. Elbistan, Nucl. Phys. B \textbf{980}, (2022) 115846

\bibitem{ppwaves} M. Elbistan, N. Dimakis, K. Andrzejewski, P. A. Horvathy, P. Kosi\'nski and P.-M. Zhang, Annals Phys. \textbf{418}, (2020) 168180

\bibitem{BFgeo} M. Elbistan, P.-M. Zhang, N. Dimakis, G. W. Gibbons and P. A. Horvathy, Phys. Rev. D \textbf{102}, (2020) 024014

\bibitem{Goenner} R. Sippel and H. Goenner, Gen. Rel. Grav. \textbf{18}, (1986) 1229

\bibitem{Maartens} R. Maartens and S. D. Maharaj, Clas. Quantum Grav. \textbf{8}, (1991) 503

\bibitem{Tupper} A. J. Keane and B. O. J. Tupper, Clas. Quantum Grav. \textbf{21}, (2004) 2037

\bibitem{Kundt} J. Ehlers and W. Kundt, ``\textit{Gravitation: An Introduction to Current Research}'', L. Witten (Ed.), Wiley, New York, pp. 49-101 (1962)

\bibitem{SthephMac} H. Stephani, et. al. ``\textit{Exact Solutions of Einstein's Field Equations}'', 2nd ed., Cambridge University Press, Cambridge (2003)

\bibitem{Steele}  J. D. Steele, ``On generalised pp waves'', http://web.maths.unsw.edu.au/˜jds/Papers/gppwaves.pdf [Retrieved 14 April 2010].

\bibitem{Coley} A. Coley, R. Milson, V. Pravda and A. Pravdova, Class. Quant. Grav. \textbf{21}, (2004) 5519

\bibitem{Penlim} R. Penrose, ``\textit{Differential Geometry and Relativity.}'' Mathematical Physics and Applied Mathematics, vol 3., M. Cahen, M. Flato (Eds.) Springer, Dordrecht (1976)

\bibitem{Ortin} T. Ortin, ``\textit{Gravity and Strings}'', Cambridge University Press, Cambridge, New York, Melbourne (2004)

\bibitem{einbein} L. Brink, P. Di Vecchia and P. Howe, Nucl. Phys. B \textbf{118}, (1977) 76

\bibitem{Dirac} P. A. M. Dirac, Canad. J. Math \textbf{2}, (1950) 129

\bibitem{Sund} K. Sundermeyer, ``\textit{Constrained Dynamics}", Springer-Verlag, Berlin, Heidelberg, New York, (1982)

\bibitem{AndBer} J. Anderson and P. Bergmann, Phys. Rev. \textbf{83}, (1951) 1018

\bibitem{tchris} T. Christodoulakis, N. Dimakis and Petros A. Terzis, J. Phys. A: Math. Theor. \textbf{47}, (2014) 095202

\bibitem{Livine1} J. B. Achour and E. R. Livine, JHEP \textbf{2020}, (2020) 67

\bibitem{Livine2} M. Geiller, E. R. Livine and F. Sartini, arXiv preprint: 2205.02615 [gr-qc] (2022)

\bibitem{Olver} P. J. Olver, ``\textit{Applications of Lie Groups to Differential Equations}", second ed., Springer, Berlin, (2000)

\bibitem{Bekenstein} J. D. Bekenstein, Phys. Rev. D \textbf{48}, (1993) 3641

\bibitem{Lobo} I. P. Lobo and G. G. Carvalho, IJGMMP \textbf{16}, (2019) 1950180

\bibitem{BenAchour} J. Ben Achour, D. Langlois and K. Noui, Phys. Rev. D \textbf{93}, (2016) 124005

\bibitem{disf1} N. Deruelle and J. Rua, JCAP 09, (2014) 002

\bibitem{disf2} E. Bittencourt, I. P. Lobo, G. G. Carvalho, Class. Quantum Grav. \textbf{32}, (2015) 185016

\bibitem{disf3} G. G. Carvalho, I. P. Lobo, E. Bittencourt, Phys. Rev. D \textbf{93}, (2016) 044005

\bibitem{disf4} M. Hohmann, Universe \textbf{5}, (2019) 167

\bibitem{disf5} E. Bittencourt, G. G. Carvalho, I. P. Lobo, L. Santana, Eur. Phys. J. C \textbf{80}, (2020) 265

\bibitem{disf6} S. Chowdhury, K. Pal, K. Pal and T. Sarkar, Eur. Phys. J. C \textbf{81}, (2021) 946

\bibitem{Bogo1} G. Y. Bogoslovsky, Il Nuovo Cimento \textbf{40 B}, (1977) 99

\bibitem{Bogo2} G. Y. Bogoslovsky, Il Nuovo Cimento \textbf{40 B}, (1977) 116

\bibitem{GiGoPo1} G. W. Gibbons, J. Gomis and C. N. Pope, Phys. Rev. D \textbf{76}, (2007) 081701(R)

\bibitem{VSL} A. G. Cohen and S. L. Glashow, Phys. Rev. Lett. \textbf{97}, (2006) 021601

\bibitem{Bogo3} G. Y. Bogoslovsky, Fortschr. Phys. \textbf{42}, (1994) 143

\bibitem{Stavrinos} A. P. Kouretsis, M. Stathakopoulos and P. C. Stavrinos, Phys. Rev. D \textbf{79}, (2009) 104011

\bibitem{Rox} I. W. Roxburgh,  Gen. Rel. Grav. \textbf{23}, (1991) 1071

\bibitem{Baobook} D. Bao, S.-S. Chern and Z. Shen, ``\textit{An Introduction to Riemann-Finsler Geometry}'', Springer-Verlag, New York, (2000)

\bibitem{Dimletter} N. Dimakis, Phys. Rev. D \textbf{103}, (2021) L071701

\bibitem{Car3} M. Cariglia, T. Houri, P. Krtou\v{s} and D. Kubiz\v{n}\'ak, Eur. Phys. J. C  \textbf{78}, (2018) 661

\bibitem{Pauli} W. Pauli, ``\textit{\"Uber das Wasserstoffspektrum vom Standpunkt der neuen Quantenmechanik}'', Z. Phys. \textbf{36}, (1926) 336 [Translation in: ``\textit{Sources of Quantum Mechanics}'' B. L. van der Waerden (Ed.), Dover Pub;ications, New York, (1967)]

\bibitem{Jala} S. Jalalzadeh, T. Rostami and P. V. Moniz, Int. J. Mod. Phys. D \textbf{25}, (2016) 1630009

\bibitem{QM1} S. Benenti, C. Chanu and G. Rastelli, J. Math. Phys. \textbf{43}, (2002) 5223

\bibitem{QM2} C. Duval and G. valent, J. Math. Phys. \textbf{46}, (2005) 053516

\bibitem{QM3} J.-P. Michel, F. Radoux and J. \v{S}ilhan, SIGMA \textbf{10}, (2014) 016

\bibitem{Plylast} S. M. Klishevich and  M. S. Plyushchay, Nucl. Phys. B \textbf{606}, (2001) 583

\end{thebibliography}
\end{document}